\documentclass[10pt]{article}
\usepackage[utf8]{inputenc}
\usepackage[T1]{fontenc}
\usepackage{microtype}
\usepackage[a4paper,top=20mm,bottom=25mm,left=15mm,right=15mm]{geometry}
\usepackage{xcolor}
\usepackage[english]{babel}
\usepackage{amsmath,amsfonts,amssymb}
\usepackage[superscript,biblabel,nomove]{cite}
\usepackage{subcaption}
\usepackage{caption}
\usepackage{booktabs}
\usepackage{graphicx}
\usepackage{placeins}
\usepackage{float}
\usepackage{tabularx}
\usepackage[colorlinks=true, allcolors=blue]{hyperref}

\graphicspath{{figures/}{./}} 
\newcommand{\MODELNAME}{AMARO v2.0}

\title{A unified coarse-grained neural network potential for the dynamics of biomolecular complexes}
\author{%
\parbox{0.95\textwidth}{\centering
Antonio Mirarchi$^{1}$, Stephen E. Farr$^{3}$, Adria Perez$^{3}$ and Gianni De Fabritiis$^{1,2,3,*}$\\[0.75em]
\small $^{1}$Computational Science Laboratory, Universitat Pompeu Fabra, Barcelona Biomedical Research Park, Carrer Dr.\ Aiguader 88, Barcelona 08003, Spain\\
\small $^{2}$Institució Catalana de Recerca i Estudis Avançats, Passeig Lluís Companys 23, Barcelona 08010, Spain\\
\small $^{3}$Acellera Labs, Doctor Trueta 183, Barcelona 08005, Spain\\[0.5em]
\small $^{*}$Correspondence: g.defabritiis@gmail.com
}}
\date{}

\begin{document}
\twocolumn[
\begin{@twocolumnfalse}
\maketitle
\begin{abstract}

\noindent The behavior of biomolecular complexes is governed by chemically specific interactions and often involves rare events that atomistic molecular dynamics struggle to capture within accessible timescales. Here we introduce \MODELNAME{}, a coarse-grained neural network potential that retains all heavy atoms and treats solvent implicitly. Hydrogen-aware bead types and total-charge conditioning provide a shared representation for proteins, drug-like molecules, and their complexes, including ternary protein--protein--ligand systems. The model preserves folded proteins and captures key features of protein conformational free-energy landscapes. Across 262 complexes spanning 11 targets, the learned potential retains most ligand poses, although quaternary structure reorientation is observed for the HIF2A ternary complex. In targeted case studies, \MODELNAME{} distinguishes the two protonation states of trypsin--benzamidine and reaches the crystal binding pose in 3.3-fold more trajectories than all-atom molecular mechanics at a fixed estimated GPU-time budget. The same framework also recovers a reference binding pose in TYK2 and supports binding free-energy estimation for both trypsin--benzamidine and TYK2--EJM31, although the resulting values are more energetically favorable than the experimental references. SHP2 and CDK8 systems expose remaining limitations in binding-pose recovery and site specificity, respectively. Together, these results support \MODELNAME{} as a shared heavy-atom potential for biomolecular complex dynamics.

\end{abstract}
\vspace{0.75em}
\end{@twocolumnfalse}
]

\section{Introduction}
Molecular recognition regulates enzyme activity and receptor signaling and underlies the action of many drugs, making it central to molecular biology and structure-based drug design \cite{pang2017rate,bleicher2003hit,jorgensen2004many}. Static structures show bound poses, and thermodynamic measurements quantify binding stability, but neither captures the binding process itself. A ligand may first diffuse around the protein and form transient surface contacts, while entry into the pocket can involve changes in protein conformation and intermediate ligand poses \cite{shan2011how,dror2011pathway}. Atomistic molecular dynamics (MD) can describe these interactions over time and, with sufficient sampling, relate binding mechanisms to thermodynamic and kinetic observables \cite{dror2012biomolecular,devivo2016role}. However, binding events are often rare on accessible timescales, so reconstructing complete pathways may require many trajectories and tens of microseconds of aggregate sampling \cite{buch2011complete,plattner2015protein}. Reduced models must therefore improve sampling without sacrificing the chemical detail required to describe molecular recognition and binding-relevant interactions \cite{kmiecik2016coarse, wagner2016representability}.

Classical molecular mechanics (MM) describes biomolecular interactions through predefined functional forms whose parameters are typically fitted to electronic-structure calculations and experimental observables \cite{mackerell2004empirical,de2024machine}. In contrast, machine-learned interatomic potentials (MLIPs), including neural network potentials (NNPs), learn flexible many-body representations from reference energies, forces, or both \cite{behler2007generalized,smith2017ani,unke2021machine}. Many modern NNPs use equivariant architectures that respect molecular symmetries \cite{simeon2023tensornet,musaelian2023allegro,batatia2022mace}, while conditioning on molecular charge can distinguish geometrically similar configurations in different charge states \cite{ko2021charge,simeon2025broadening,anstine2025aimnet2}. Fully atomistic NNPs, however, remain costly for large biomolecular systems, particularly when explicit solvent further increases the number of atoms included in each force evaluation \cite{eastman2026accuracy}. Hybrid NNP/MM approaches can reduce this cost by applying the learned potential to the ligand while treating the protein and solvent classically, but this requires separate models for the molecular components and an explicit treatment of their cross-interactions \cite{lahey2020simulating,sabaneszariquiey2024enhancing, farr2026evaluating}.

Coarse-graining provides a complementary route by integrating out selected degrees of freedom, reducing the number of interaction sites and potentially smoothing the effective energy landscape \cite{kmiecik2016coarse}. For example, the general-purpose Martini force field extends this strategy across chemically diverse systems \cite{souza2021martini3} and has enabled unbiased simulations of protein--ligand binding over millisecond aggregate timescales \cite{souza2020proteinligand}. These gains come with an inherent trade-off: the mapping determines which molecular information is retained and therefore balances computational efficiency against representational detail \cite{wagner2016representability,rudzinski2014investigation}. Machine-learned coarse-grained potentials offer a systematic way to learn the resulting many-body effective interactions directly from atomistic data \cite{noid2008multiscale,wang2019cgnet,husic2020cg}. However, their construction through force matching relies on noisy projected forces and can require large atomistic training datasets \cite{kramer2023statistically,durumeric2026learning}. Recent coarse-grained protein NNPs range from single-C$\alpha$ representations to mappings that retain the backbone heavy atoms and C$\beta$, and have reproduced folding and conformational free-energy landscapes across multiple proteins \cite{majewski2023machine,charron2025navigating,durumeric2026learning}.
These models remain largely protein-centric, and some rely on fixed physical priors to prevent sampling of nonphysical configurations \cite{charron2025navigating,durumeric2026learning}. More importantly for protein--ligand applications, their protein-specific mappings do not naturally extend to chemically diverse small molecules, requiring a separate ligand representation and a compatible description of protein--ligand cross-interactions.

AMARO v1.0 introduced the non-hydrogen (\textit{noh}) mapping for proteins, in which each heavy atom is retained as a coarse-grained site, and its attached hydrogens are encoded in the local bead type \cite{mirarchi2024amaro}. The resulting NNP learns many-body effective interactions from mapped forces without fixed energetic priors or structural restraints. Building on this foundation, we introduce \MODELNAME{}, which applies the same mapping across proteins and drug-like molecules while implicitly treating solvent and counterions. Here, neural charge-equilibration (NQE) blocks \cite{anstine2025aimnet2,farr2026aceff} condition the learned representation on the total system charge without introducing an explicit Coulomb energy term.

We trained \MODELNAME{} sequentially on quantum-mechanical small-molecule data, protein conformations, and protein--ligand complexes. We first evaluated protein conformational behavior and small-molecule stability, followed by protonation-sensitive interactions and binding-pose recovery in trypsin--benzamidine. We then assessed short-timescale structural retention across 262 protein--ligand complexes spanning 11 targets, before examining binding-region exploration from unbound configurations in TYK2, SHP2, and CDK8.

\section{Results}

\MODELNAME{} uses TensorNet2 \cite{farr2026aceff}, an equivariant neural network based on Cartesian tensor representations \cite{simeon2023tensornet}, with neural charge-equilibration (NQE) blocks for total-charge conditioning \cite{anstine2025aimnet2}. These blocks predict site-wise scalar features and redistribution weights, then use the supplied total system charge to normalize each feature channel across the represented graph. These normalized features enter subsequent message-passing interactions. In \MODELNAME{}, these quantities serve as latent features: they are learned through the force-matching objective without partial-charge or dipole-moment supervision and are not used to compute an explicit Coulomb energy. The network predicts a scalar effective potential, and forces are obtained as its negative gradient.

The \textit{noh} mapping retains each heavy atom as a coarse-grained site with its atomic mass and encodes the number of attached hydrogens in its bead type. The original AMARO representation contains 12 protein bead types: CH0--CH3, NH0--NH3, OH0--OH1, and SH0--SH1 \cite{mirarchi2024amaro}. We extended the representation with B, F, Si, P, Cl, Br, I, Na, K, Ca, Mg, Co, Cu, and Zn, resulting in 26 bead types spanning 18 elements. Solvent and counterions are omitted from the mapped systems. For small molecules, we obtained the total formal charge from the molecular metadata; for proteins and protein--ligand complexes, we assigned the total charge from residue protonation states at pH~7.45 using HTMD \cite{doerr2016htmd}.

We constructed the force-matching targets using the basic force-aggregation scheme of Kr\"amer et al. \cite{kramer2023statistically}, as in AMARO v1.0 \cite{mirarchi2024amaro}. For the $i$th \textit{noh} bead,

\begin{equation}
    \mathbf{F}_i
    =
    \mathbf{f}_i
    +
    \sum_{j \in H(i)} \mathbf{f}_j ,
\end{equation}
where $\mathbf{f}_i$ is the force on the retained heavy atom and $H(i)$ is the set of hydrogen atoms bonded to it.

\begin{figure*}[!t]
    \centering
    \includegraphics[width=\textwidth]{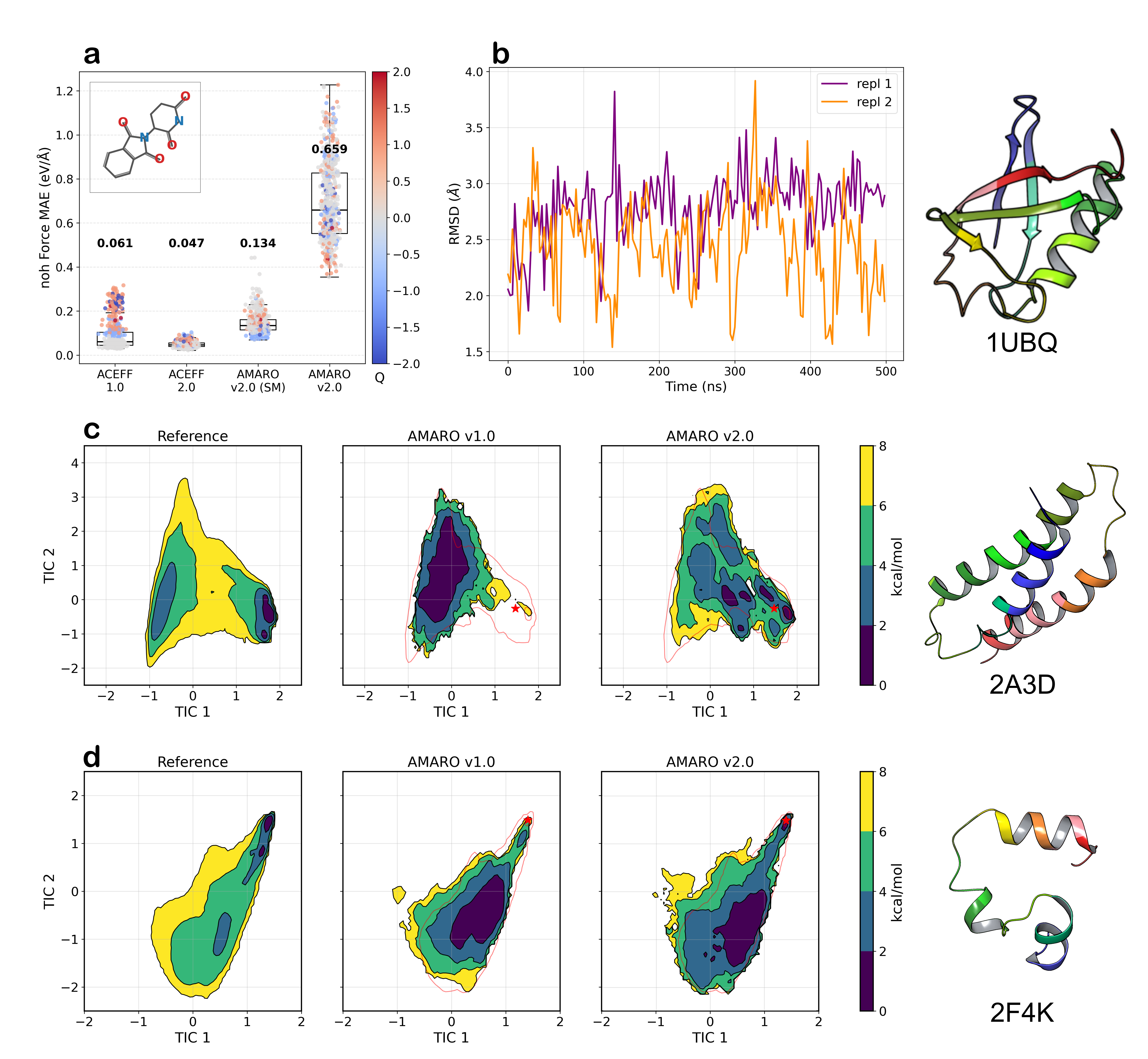}
    \caption{Validation across drug-like molecules and folded proteins. (a) Mean absolute error (MAE) of forces per conformer for 650 drug-like conformers evaluated against \textit{noh} reference forces obtained by force aggregation. Points are colored by total formal charge, boxes summarize the distributions, and bold values report the median errors in eV/\AA{}. For AceFF, each predicted hydrogen force was added to that of its bonded heavy atom before evaluation. (b) \textit{noh} root-mean-square deviation (RMSD) relative to PDB 1UBQ for two independent 500~ns \MODELNAME{} simulations at 300~K, with the reference structure shown on the right. (c,d) Conformational free-energy landscapes of $\alpha$3D and villin for the all-atom reference, AMARO v1.0, and \MODELNAME{}. Relative free energies were obtained from Markov state model (MSM) equilibrium probabilities, setting each minimum to zero. Red outlines show the reference contours; stars mark the experimental reference structures which are shown on the right.}
    \label{fig:overall_validation}
\end{figure*}

\subsection{Validation studies}

\subsubsection{Folded-state dynamics of ubiquitin}
Ubiquitin is a 76-residue protein with a compact crystal structure (PDB ID: 1UBQ) comprising a five-stranded $\beta$-sheet, an $\alpha$-helix, and a short $3_{10}$-helix \cite{vijay1987structure}. Its disulfide-free fold and high thermal stability make it a useful system for assessing folded-state stability \cite{wintrode1994thermodynamics,herberhold2002temperature}.

We ran two independent 500~ns simulations of folded ubiquitin at 300~K using its \textit{noh} representation, which contains 602 beads and has a net charge of zero at pH~7.45. In both replicas, the \textit{noh}-RMSD remained predominantly between 1.5 and 3.0~\AA{}, with only brief excursions and no progressive structural drift (Figure~\ref{fig:overall_validation}b).

\subsubsection{Protein conformational landscapes}
We examined the conformational distributions of two fast-folding proteins \cite{lindorff2011fast}, $\alpha$3D and villin (Figure~\ref{fig:overall_validation}c,d) using the same time-lagged independent component analysis (TICA) and Markov state model (MSM) workflow as in AMARO v1.0. Each protein--model ensemble comprised 32 simulations initialized from conformations selected uniformly across the two-dimensional all-atom reference TICA free-energy surface (FES).

$\alpha$3D, a 73-residue three-helix bundle, showed the largest deviation from the all-atom conformational landscape among the four fast-folding proteins evaluated with the first AMARO model \cite{mirarchi2024amaro}. In the common TICA projection, the predicted landscape contains a broad low-free-energy region near the center, whereas the reference global minimum lies on the right. \MODELNAME{} reduces this central-region oversampling and more closely reproduces the all-atom distribution, with increased probability in the right-hand basin and along the lower-TIC2 branch.

Villin provides a complementary test because AMARO v1.0 already captures the overall shape of its reference free-energy landscape. Both AMARO models sample the branch containing the crystal structure, but retain a broad low-free-energy basin at lower TIC values that is less favorable in the all-atom reference. Compared with AMARO v1.0, \MODELNAME{} shows a more pronounced low-free-energy region near the crystal structure, while differences elsewhere in the landscape remain.

\subsubsection{Force prediction for drug-like molecules}

To assess force prediction for drug-like molecules, we used the AceFF v2.0 benchmark \cite{farr2026aceff}, which comprises 650 conformers from the Schrödinger ligand set \cite{ross2023maximal} and includes molecules larger than those in the small-molecule training data. We calculated reference forces at the same $\omega$B97M-V/def2-TZVPPD level used for the small-molecule training data of both AceFF and \MODELNAME{}, then mapped them to the \textit{noh} representation using \textit{basic} force aggregation \cite{kramer2023statistically}. For AceFF, the predicted all-atom forces were mapped equivalently by adding each hydrogen force to that of its bonded heavy atom before calculating the mean absolute error (MAE).

Our evaluation reproduces the improvement from AceFF v1.0 to v2.0 reported in the original benchmark, with median MAEs of 0.061 and 0.047~eV/\AA{}, respectively (Figure~\ref{fig:overall_validation}a).
The small-molecule checkpoint, \MODELNAME{} (SM), yields a median MAE of 0.134~eV/\AA{}, about threefold higher than AceFF v2.0. Its errors show no clear dependence on total molecular charge.
The final checkpoint yields a median MAE of 0.659~eV/\AA{}, approximately fivefold higher than \MODELNAME{} (SM). Because the two checkpoints share the same architecture and representation, this loss in small-molecule force accuracy suggests a trade-off in the subsequent protein and protein--ligand training stages.

\subsubsection{Small-molecule dynamics and local geometry}
To complement the force benchmark, we examined the dynamics of three ligands selected from distinct subsets of the Schrödinger public protein--ligand binding free-energy benchmark \cite{ross2023maximal}: JNK1 from the JACS/FEP+ R-group subset, TNKS2 from the Merck subset, and JAK1 from the FEP+ charge-change subset, with formal charges of \(-1\), 0, and \(+1\), respectively. For each potential and time step, we ran three independent 5~ns simulations of each isolated molecule at 300~K, without explicit solvent, and calculated the heavy-atom RMSD relative to the starting configuration (Supplementary Figure~\ref{sup_fig:dynamic_small_mol}).
The final checkpoint completed all 27 simulations, including every 4~fs run, despite its larger force MAE. We then compared heavy-atom bond lengths, bond angles, out-of-plane angles, and 1--4 distances with matched AceFF v2.0 simulations (Supplementary Figure~\ref{sup_fig:small_mol_geometry}). Across the analyzed conditions, \MODELNAME{} (SM) more closely reproduced the AceFF heavy-atom local-geometry distributions than the final checkpoint. At 1~fs, median bond-length $W_1$ values were 0.0058~\AA{} for the AceFF baseline, 0.0143~\AA{} for \MODELNAME{}, and 0.0060~\AA{} for \MODELNAME{} (SM); the corresponding bond-angle values were 0.54$^\circ$, 1.64$^\circ$, and 0.62$^\circ$. Thus, relative to matched AceFF v2.0 simulations, \MODELNAME{} (SM) more closely reproduced the sampled heavy-atom local-geometry distributions, whereas the final checkpoint completed all simulations without numerical instability, including at 4~fs.

\subsection{Trypsin--benzamidine case study}
We used bovine $\beta$-trypsin--benzamidine systems based on PDB structures 1BTY and 3PTB as a detailed protein--ligand case study of the full coarse-grained NNP framework. Previous atomistic simulations reconstructed the complete binding process for this well-studied system \cite{buch2011complete,plattner2015protein}, while later enhanced-sampling studies examined alternative unbinding pathways \cite{deeks2023free}. We therefore evaluated protonation-state sensitivity, bound-state contacts, and binding from unbound configurations, followed by MSM analyses of the binding landscape and thermodynamics.

\begin{figure*}[t]
    \centering
    \includegraphics[width=\textwidth]{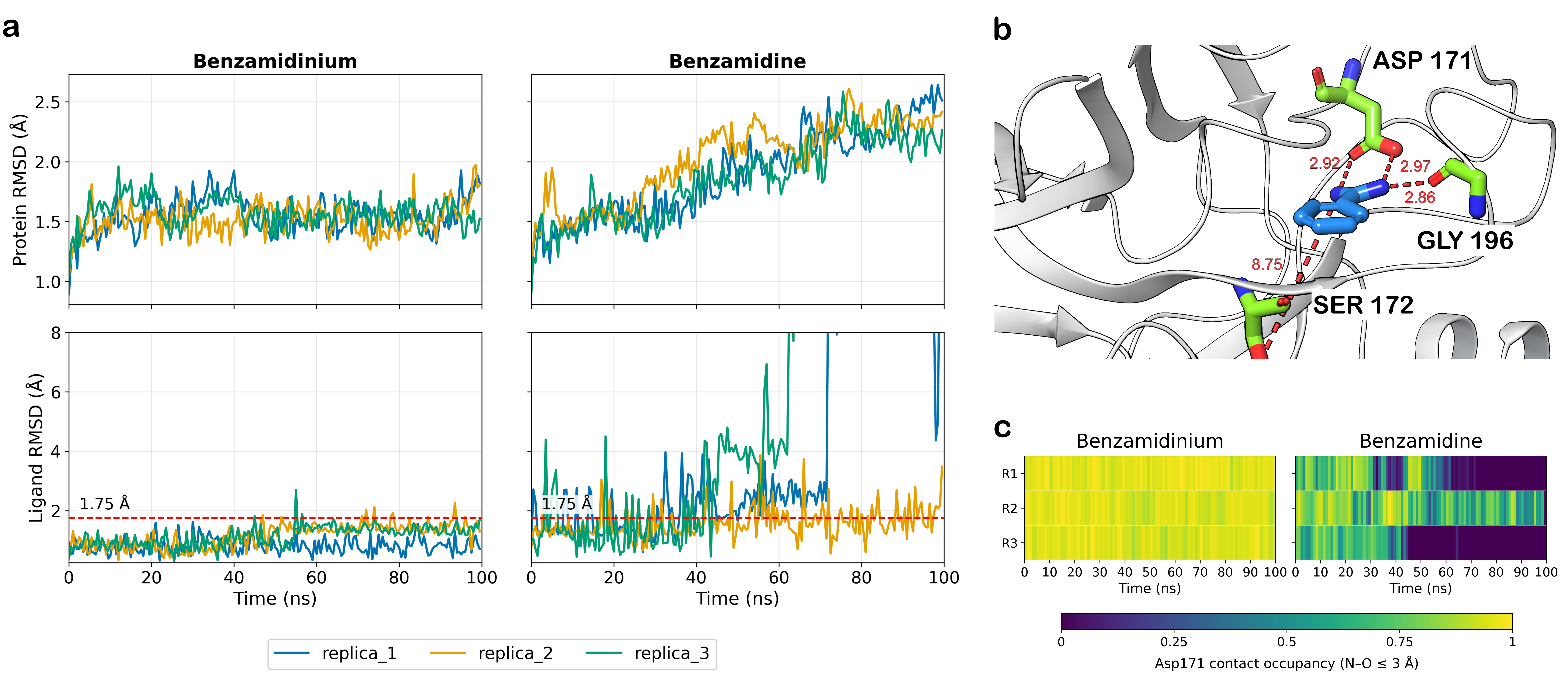}
    \caption{Protonation-state sensitivity in the trypsin S1 pocket. (a) Protein (top) and ligand (bottom) RMSDs for three independent 100~ns replicas of each protonation state, benzamidinium and neutral benzamidine. RMSDs were calculated after alignment to the protein C$\alpha$ atoms of PDB 1BTY. Red dashed lines mark a ligand RMSD of 1.75~\AA{}. (b) Binding-site geometry of benzamidinium near Asp171, Ser172, and Gly196; dashed lines indicate selected distances in \AA{}. (c) Asp171 contact occupancy for each replica (R1--R3). Each heatmap cell reports the fraction of frames in a 1~ns window for which at least one of the four distances N1--OD1, N1--OD2, N2--OD1, or N2--OD2 is $\leq3.0$~\AA{}, where N1/N2 are the ligand nitrogens and OD1/OD2 are the Asp171 carboxylate oxygens. Simultaneous contact by both ligand nitrogens is not required. The 3.0~\AA{} cutoff captures close N--O contacts consistent with the Asp171 interaction distances shown in panel b.}
    \label{fig:benzamidine_rmsd}
\end{figure*}

\subsubsection{Protonation-state sensitivity}
Benzamidine has a reported p$K_\mathrm{a}$ of 11.6 and is therefore predominantly protonated as benzamidinium at pH~7.45 \cite{talhout2001thermodynamic}. In the crystallographic S1 pocket, the resulting amidinium group forms a key interaction with Asp171 (Figure~\ref{fig:benzamidine_rmsd}b), making this system a focused test of protonation-sensitive binding behavior. We therefore simulated the protonated and neutral forms in three independent 100~ns replicas each, initialized from the crystallographic pose and using the corresponding ligand bead types and total solute charges of \(+9\) for benzamidinium and \(+8\) for neutral benzamidine. No structural or positional restraints were applied. The protonated ligand remains close to the crystal pose in all three replicas, with ligand RMSD mostly below 1.75~\AA{} and protein RMSD near 1--2~\AA{} (Figure~\ref{fig:benzamidine_rmsd}a). In contrast, neutral benzamidine shows larger ligand RMSD fluctuations, with large displacements from the crystal pose in replicas 1 and 3 during the second half of the simulations, whereas replica 2 shows no clear unbinding over 100~ns. Protein RMSDs remain below approximately 3~\AA{} in both protonation states, indicating that the larger displacements primarily involve the ligand.

To relate these differences in pose retention to S1-pocket interactions, we quantified the occupancy of contacts between the ligand nitrogens and the Asp171 carboxylate oxygens (Figure~\ref{fig:benzamidine_rmsd}c). Benzamidinium maintains high contact occupancy across all three replicas throughout the 100~ns simulations, consistent with the expected salt bridge between the cationic ligand and the anionic side chain \cite{talhout2001thermodynamic,kurczab2018salt}. Neutral benzamidine instead shows lower and more variable contact occupancy, with sustained contact loss in replicas 1 and 3 during the latter part of the simulations. In replica 2, reduced and fluctuating Asp171 contact occupancy shows that this interaction is less persistent even while the ligand remains near the crystal pose. Together, the RMSD and contact patterns support greater bound-pose stability of benzamidinium in the S1 pocket over the 100~ns observation window.

\subsubsection{Binding reconstruction from unbound configurations}

We examined binding from 100 unbound configurations selected from the deposited all-atom dataset \cite{plattner2015protein, buch2011complete}, with initial ligand--pocket separations ranging from approximately 15 to 35~\AA{}. The 3PTB-based system retained the calcium ion and had a total solute charge of $+9$. Each mapped configuration was simulated for 50~ns with protein C$\alpha$ restraints and a flat-bottom potential limiting ligand diffusion away, yielding 5.0~$\mu$s of aggregate \MODELNAME{} sampling. Within the 50-ns observation window, pose recovery was observed in 42\% of the \MODELNAME{} trajectories (42/100; 95\% Wilson CI, 32.8--51.8\%), compared with 4\% of the corresponding all-atom trajectories (4/100; 95\% CI, 1.6--9.8\%; Supplementary Figure~\ref{sup_fig:tryps_pose_recovery}). The recovery fraction was therefore 38 percentage points higher for \MODELNAME{} (95\% paired-bootstrap CI, 28--48 percentage points).

\begin{figure*}[t]
    \centering
    \includegraphics[width=\textwidth]{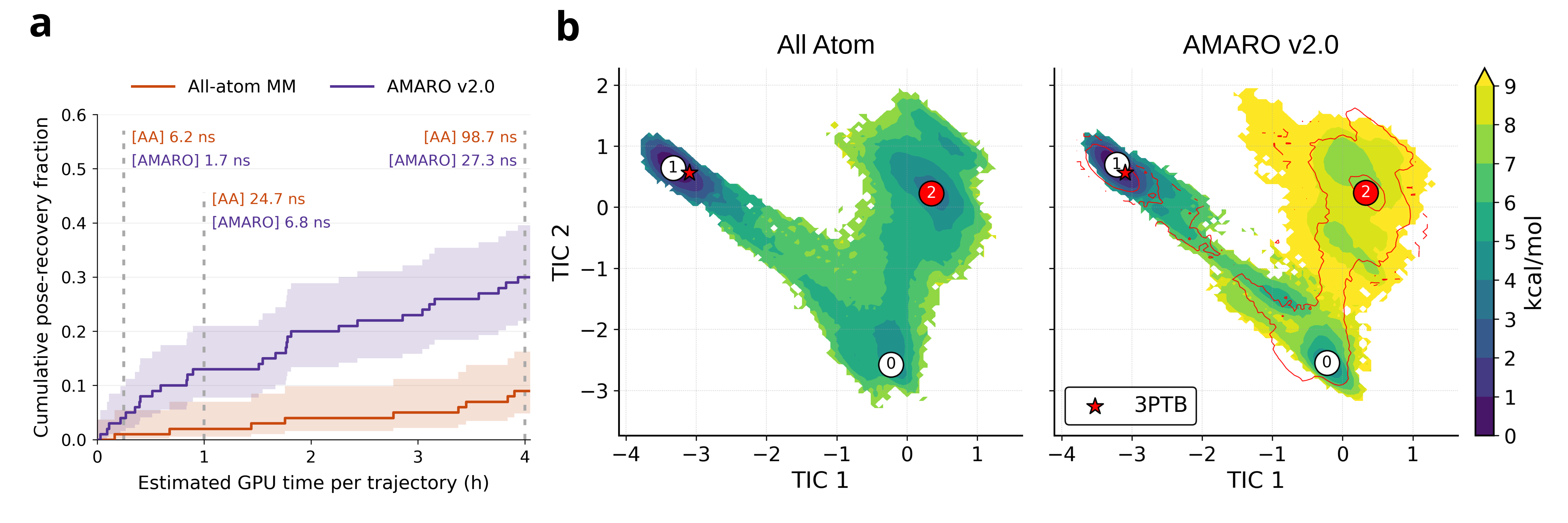}
    \caption{Trypsin--benzamidine pose recovery versus GPU time and free-energy landscapes. (a) Cumulative fraction of 100 matched trajectories per model reaching the 3PTB pose at least once, plotted against estimated GPU time per trajectory. Bound pose recovery was defined by a ligand RMSD of 2.5~\AA{} or less after protein C$\alpha$ alignment. GPU time was calculated from median throughputs of 592~ns/day for all-atom MM and 164~ns/day for \MODELNAME{}. Solid step curves show empirical cumulative fractions, and shaded bands show pointwise 95\% Wilson confidence intervals. Vertical dashed lines mark selected computational budgets; annotations give the corresponding simulated durations for all-atom MM (AA) and \MODELNAME{} (AMARO). (b) Independently estimated all-atom (left) and \MODELNAME{} (right) MSM landscapes in a common TICA space fitted to all-atom ligand--residue COM distances. Both models contain three final macrostates. Numbered circles mark macrostate centers: M1 is the bound state, M2 (red circle) is the dedicated bulk state, and M0 is the remaining nonbulk state. Red stars mark the projected 3PTB structure, and red contours show the all-atom landscape. Both landscapes share the kcal/mol scale.}
    \label{fig:tryps_binding_msm}
\end{figure*}

Moreover, the higher fraction of trajectories reaching the bound conformation persisted at a fixed computational budget (Figure~\ref{fig:tryps_binding_msm}a): within an estimated 4~GPU hours per trajectory, 30\% of \MODELNAME{} trajectories had reached the reference pose at least once (30/100; 95\% Wilson CI, 21.9--39.6\%), compared with 9\% of all-atom trajectories (9/100; 95\% CI, 4.8--16.2\%), a 3.3-fold difference. The recovery fraction was 21 percentage points higher for \MODELNAME{} (95\% paired-bootstrap CI, 12--30 percentage points). This 4-hour GPU budget corresponds to 27.3~ns of \MODELNAME{} simulation and 98.7~ns of all-atom simulation per trajectory.

We next characterized the binding landscapes in a common low-dimensional representation by projecting the \MODELNAME{} trajectories onto TICA coordinates fitted to all-atom ligand--residue center-of-mass (COM) distances and independently discretizing the two ensembles (Figure~\ref{fig:tryps_binding_msm}b). For the thermodynamic analysis, frames whose ligand COM was farther than 12~\AA{} from every protein-residue COM were reassigned to a distinct new bulk microstate for each mixed microstate, while nonbulk frames retained their original labels. After fitting the MSM and PCCA+ partition, all bulk microstates were grouped into a dedicated bulk macrostate without changing the fitted microstate transition matrix. Both models resolve three final macrostates in corresponding regions of TICA space: M1 is the reference-pose-rich bound state, M2 is the dedicated bulk state, and M0 contains the remaining nonbulk configurations (Supplementary Figure~\ref{sup_fig:tryps_macrostate_composition}). We calculated binding free energies from the stationary bulk and bound populations using the effective volume associated with the bulk definition. Across bulk cutoffs of 8--20~\AA{}, the trypsin estimates varied by 0.06~kcal/mol or less for both models, whereas smaller cutoffs gave less favorable binding free energies (Supplementary Figure~\ref{sup_fig:dg_vs_bulk_cutoff}).

Table~\ref{tab:binding_free_energies} compares the binding free-energy estimates with experimental reference values. For the trypsin complex, the all-atom estimate was $-5.80\pm0.39$~kcal/mol, compared with $-6.20$~kcal/mol experimentally, whereas \MODELNAME{} yielded $-11.59\pm0.74$~kcal/mol. The \MODELNAME{} estimate was therefore 5.39~kcal/mol more favorable than the experimental value; uncertainties represent sample standard deviations of 100 bootstrap estimates. The higher relative free energy of the bulk region in the \MODELNAME{} landscape (Figure~\ref{fig:tryps_binding_msm}b) is consistent with its more negative binding free-energy estimate, reflecting a stronger population bias toward the bound state. We calculated the association ($k_{\mathrm{on}}$) and dissociation ($k_{\mathrm{off}}$) rate constants using HTMD~\cite{doerr2016htmd}, employing the same reassigned-bulk source and reference-pose sink states used for the $\Delta G^\circ$ calculation, and compared our estimates with experimental data from~\cite{guillain1970trypsin}. The all-atom model yielded $k_{\mathrm{on}} = (3.98 \pm 12.11)\times10^{8}$~M$^{-1}$s$^{-1}$ and $k_{\mathrm{off}} = (3.24 \pm 2.22)\times10^{4}$~s$^{-1}$, while the \MODELNAME{} model gave $k_{\mathrm{on}} = (2.61 \pm 6.13)\times10^{9}$~M$^{-1}$s$^{-1}$ and $k_{\mathrm{off}} = 106 \pm 97$~s$^{-1}$. The experimental values are $k_{\mathrm{on}} = 2.9\times10^{7}$~M$^{-1}$s$^{-1}$ and $k_{\mathrm{off}} = 6.0\times10^{2}$~s$^{-1}$, respectively.

Both model predictions overestimate the experimental $k_{\mathrm{on}}$, although the bootstrap SD exceeds the full-data estimate for each model. For $k_{\mathrm{off}}$, the all-atom estimate is higher than the experiment, whereas the \MODELNAME{} estimate is lower. These dissociation estimates remain sensitive to limited sampling; in particular, no pose-core-to-bulk dissociation was observed for \MODELNAME{}. As a complementary characterization of the modeled kinetics, the slowest implied timescale is approximately 0.6~$\mu$s for the all-atom MSM at a 20~ns lag and 0.15~$\mu$s for \MODELNAME{} at a 10~ns lag (Supplementary Figure~\ref{sup_fig:tryps_its}).

\begin{table}[ht]
    \centering
    \caption{Binding free-energy estimates and experimental reference values. All values are in kcal/mol. Calculated values are full-data estimates at 300~K using a 12~\AA{} bulk-classification cutoff and the corresponding effective bulk volume; uncertainties are sample standard deviations of finite bootstrap estimates (100 bootstrap resamples for each trypsin model and 20 for TYK2). The trypsin--benzamidine reference is the rounded literature value \cite{plattner2015protein}, based on inhibition measurements at pH~8.15 and 288.15~K \cite{mares1965studies}. The TYK2--EJM31 reference value was derived from the reported $K_i=0.096$~$\mu$M \cite{liang2013tyk2lead,hahn2022best} using $\Delta G^\circ=RT\ln[K_i/(1~\mathrm{M})]$ at 297~K, following the benchmark convention of Wang et al.\ \cite{wang2015accurate}. A dash indicates that no all-atom estimate was calculated.}
    \label{tab:binding_free_energies}
    \small
    \setlength{\tabcolsep}{3pt}
    \begin{tabularx}{\linewidth}{@{}>{\raggedright\arraybackslash}Xccc@{}}
        \toprule
        & \multicolumn{3}{c}{$\Delta G^\circ$ (kcal/mol)} \\
        \cmidrule(l){2-4}
        System & Experiment & All-atom & \MODELNAME{} \\
        \midrule
        3PTB & $-6.20$ & $-5.80\pm0.39$ & $-11.59\pm0.74$ \\
        TYK2--\allowbreak EJM31 & $-9.54$ & --- & $-10.95\pm2.14$ \\
        \bottomrule
    \end{tabularx}
\end{table}


\subsection{Molecular recognition structural benchmark}
To assess structural retention across a broader set of protein--ligand complexes, we selected 262 cofactor-free systems from the OpenFF protein--ligand benchmark \cite{hahn2022best}, spanning 11 targets: CDK2, CDK8, cMET, HIF2A, MCL1, p38, PTP1B, SHP2, SYK, thrombin, and TYK2. We mapped, minimized, and simulated each complex for 10~ns, then evaluated protein structural stability and ligand-pose retention.

The receptors span different degrees of amino acid sequence overlap with the protein--ligand training set (Supplementary Figure~\ref{sup_fig:openff_receptor_overlap}). PTP1B and p38 have exact sequence matches, whereas cMET, MCL1, SYK, and TYK2 have counterparts with at least 95\% sequence identity and coverage. The remaining five targets fall below this threshold, with closest training-set receptors ranging from 88.1\% sequence identity for thrombin to 21.0\% for CDK8; CDK2, HIF2A, and SHP2 show intermediate identities.

\subsubsection{Protein and ligand structural retention}
\begin{figure}[t]
    \centering
    \includegraphics[width=\linewidth]{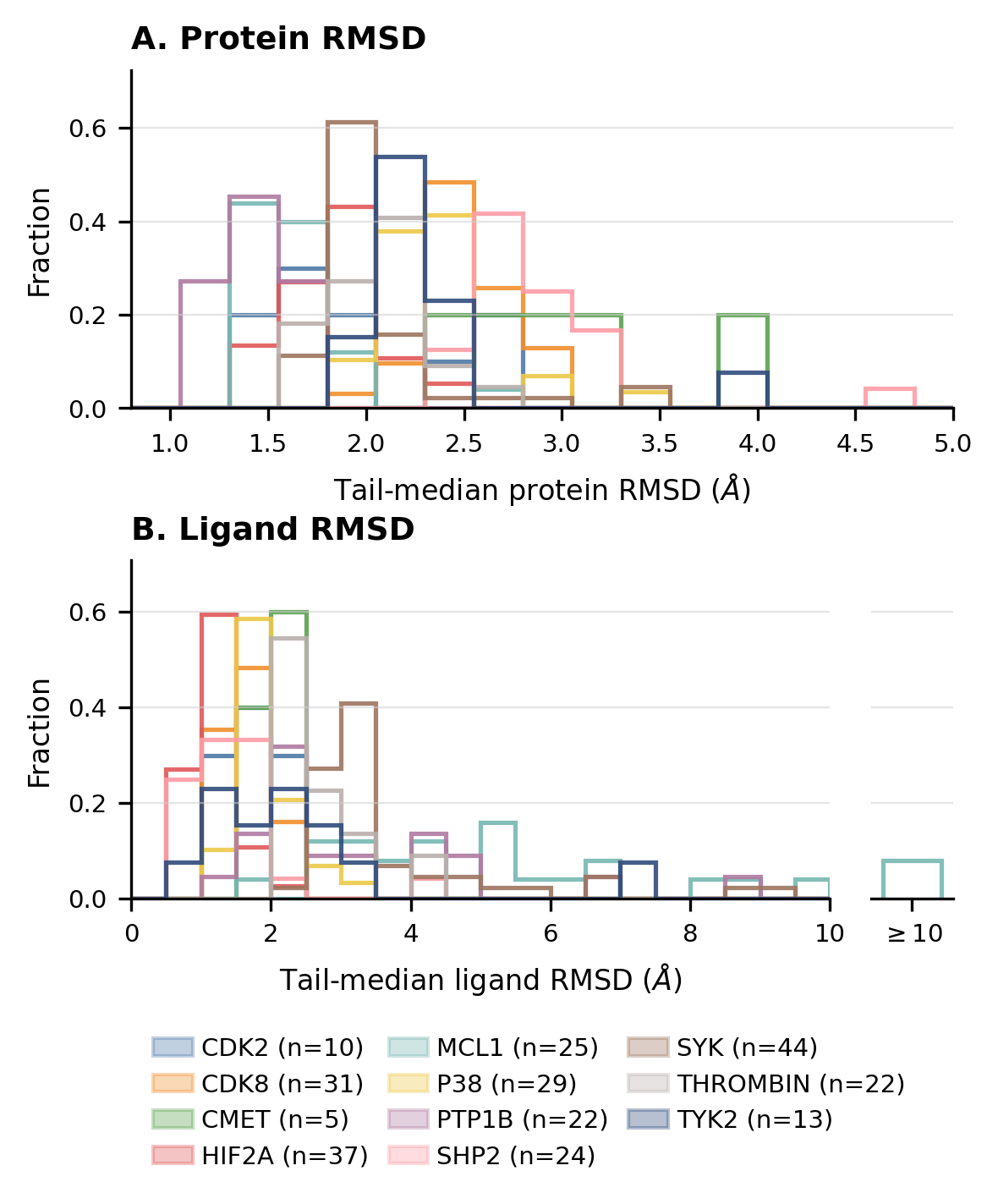}
    \caption{Target-wise summaries across the OpenFF benchmark subset. Distribution of the tail-median protein C$\alpha$-RMSD (a) and protein-aligned ligand RMSD (b), calculated over the final 25\% of each 10~ns trajectory. For heterodimers, protein alignment and RMSD calculations were restricted to the ligand-binding monomer. Histograms are normalized within each target; \(n\) indicates the number of complexes. Ligand RMSDs $\geq10$~\AA{} are grouped in a separate overflow bin.}
    \label{fig:openff_rmsd_distributions}
\end{figure}

We quantified protein structural retention using the median C$\alpha$-RMSD over the final 25\% of each trajectory (Figure~\ref{fig:openff_rmsd_distributions}a). For the heterodimeric CDK2, CDK8, and HIF2A systems, we restricted alignment and RMSD calculations to the ligand-binding monomer, separating its internal conformational stability from the partner chain's relative motion.

Across all 11 targets, most protein structures remained within approximately 1--3~\AA{} of their initial conformations, with limited structural drift over 10~ns. The ligand-binding monomers of the three heterodimers showed similarly low RMSDs; we analyze their interchain behavior separately below.

After protein alignment, we classified ligand retention using the corresponding tail-median RMSD (Figure~\ref{fig:openff_rmsd_distributions}b): retained ($<4$~\AA{}), rearranged ($4$--$6$~\AA{}), or displaced ($>6$~\AA{}). All ligands were retained for CDK2, CDK8, cMET, HIF2A, and p38. SHP2 and thrombin contained one and two rearranged ligands, respectively, while TYK2 contained one displaced ligand. MCL1, PTP1B, and SYK showed greater variability, with MCL1 exhibiting the largest fraction of altered poses (Supplementary Figure~\ref{sup_fig:openff_ligand_outcomes}). The outcome distributions showed no clear relationship with ligand or complex size (Supplementary Table~\ref{tab:openff_complex_summary}). In most targets, protein RMSD remained low when ligand poses changed, indicating that these deviations primarily reflect ligand motion rather than global complex instability. Representative MCL1 trajectories illustrate both local torsional rearrangement and larger ligand displacement (Supplementary Figure~\ref{sup_fig:mcl1_case_ligands}).

\subsubsection{Protein--protein interface retention}
\begin{figure*}[!t]
    \centering
    \includegraphics[width=\textwidth]{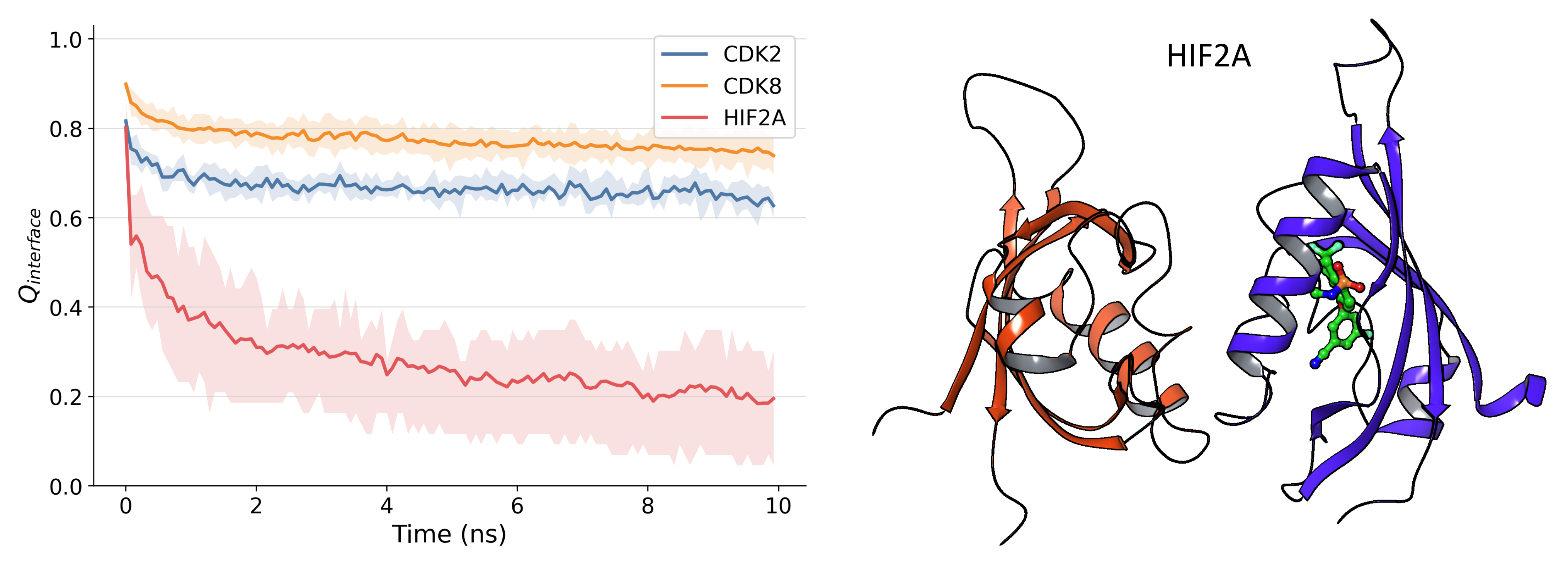}
    \caption{Protein--protein interface retention in the heterodimeric OpenFF systems. (left) Time evolution of the mean \(Q_{\mathrm{interface}}\) for CDK2, CDK8, and HIF2A; shaded regions indicate the interquartile range across complexes. (right) Representative HIF2A configuration illustrating the relative reorientation of the two chains. The ligand-binding monomer is shown in purple, the partner chain in orange, and LIG254 in green.}
    \label{fig:openff_protein_interface}
\end{figure*}

We further assessed the heterodimeric systems using the fraction of reference interchain contacts, \(Q_{\mathrm{interface}}\) (Figure~\ref{fig:openff_protein_interface}, left). CDK2 and CDK8 underwent an initial relaxation and subsequently retained approximately two-thirds of their reference contacts, with median \(Q_{\mathrm{interface}}\) values of 0.67 and 0.70 over the final 25\% of the trajectories, respectively. Both profiles plateaued, indicating substantial but incomplete interface retention over 10~ns.

HIF2A behaved differently: \(Q_{\mathrm{interface}}\) decreased throughout the simulations to a tail median of approximately 0.16. To determine whether this interface loss arises from internal conformational changes or relative motion of the two chains, we calculated the partner-chain C$\alpha$-RMSD after alignment to either the ligand-binding monomer or the partner chain itself. Across the 37 HIF2A trajectories, the corresponding tail medians were 15.0 and 1.93~\AA{}, respectively, while the median relative rigid-body rotation of the partner core was 53.8$^\circ$. Thus, interface loss primarily reflected rigid-body reorientation of the partner chain, which largely retained its internal structure (Figure~\ref{fig:openff_protein_interface}).

\subsubsection{Exploration of binding events}
\begin{figure*}[!t]
    \centering
    \includegraphics[width=0.98\textwidth]{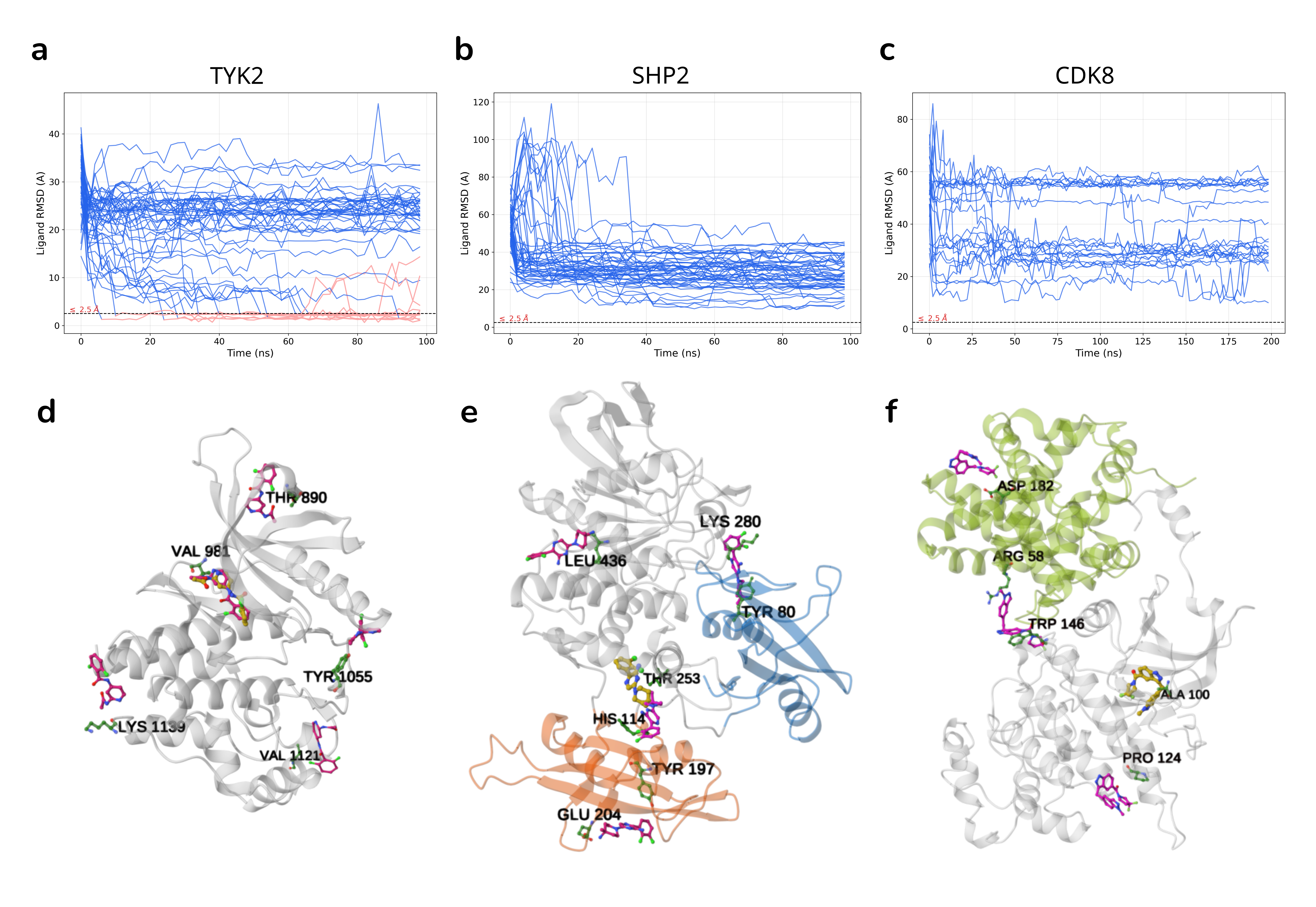}
    \caption{Binding exploration in TYK2, SHP2, and CDK8. (a--c) Protein-aligned ligand RMSD from 50 independent 100~ns trajectories for TYK2--EJM31 and SHP2--SHP099, and 25 independent 200~ns trajectories for CDK8--ligand 43. Ligand RMSDs were calculated after protein C$\alpha$ alignment to the corresponding benchmark structure. The dashed line marks 2.5~\AA{} RMSD; TYK2 trajectories reaching this threshold are shown in red. (d--f) Representative ligand configurations from the corresponding contact macrostates. Sampled configurations are shown in magenta, the reference pose in yellow, and selected nearby residues as labeled sticks. In d, TYK2 M0 overlaps the reference ATP-site pose near Val981, whereas M1--M4 occupy surface regions near Thr890, Tyr1055, Val1121, and Lys1139, respectively. In e, SHP2 M0 overlaps the N-SH2--PTP latch site near Tyr80/Lys280, M2 occupies the canonical SHP099 tunnel near His114/Thr253, and M1 and M3 represent exposed PTP and C-SH2 surface contacts near Leu436 and Tyr197/Glu204, respectively; the N-SH2, C-SH2, and PTP domains are shown in blue, orange, and light gray, respectively. In f, CDK8 M0--M2 lie near CDK8 Pro124, a region adjacent to CDK8 Trp146 and Cyclin C Arg58, and Cyclin C Asp182, respectively; CDK8 is light gray and Cyclin C is light green.}
    \label{fig:additional_binding_systems}
\end{figure*}

We took three systems from the set above and explored the extent to which binding was recovered starting from bulk.
We initiated EJM31, SHP099, and ligand 43 from multiple unbound configurations around TYK2, SHP2, and CDK8, respectively. We generated 50 independent 100~ns trajectories for TYK2--EJM31 and SHP2--SHP099 and 25 independent 200~ns trajectories for CDK8--ligand 43, yielding 5~$\mu$s of aggregate sampling per system. For each system, we used TICA and MSM analysis to identify recurrent ligand-contact macrostates and summarized their structural organization alongside the ligand RMSD traces in Figure~\ref{fig:additional_binding_systems} and Supplementary Table~\ref{sup_tab:additional_binding_macrostates}.

For TYK2, 14 of the 50 trajectories reached an EJM31 RMSD lower than 2.5~\AA{} from the reference pose. The macrostate enriched in these low-RMSD configurations was centered in the ATP cleft, and its representative structure placed the pyridine nitrogen near the Val981 backbone NH, consistent with the hinge-binding geometry proposed for this pyridine scaffold \cite{liang2013tyk2lead}. By contrast, the remaining macrostates represent contacts at other regions of the TYK2 surface (Figure~\ref{fig:additional_binding_systems}).

As for trypsin--benzamidine, the TYK2 binding free-energy estimate of $-10.95\pm2.14$~kcal/mol was more energetically favored than the experimental value of $-9.54$~kcal/mol. The estimate shifted overall toward more negative values as the bulk cutoff increased over 10--15~\AA{} (Supplementary Figure~\ref{sup_fig:dg_vs_bulk_cutoff}); state populations and sampling diagnostics are reported in Supplementary Table~\ref{sup_tab:tyk2_equilibrium_diagnostics}.

In the autoinhibited conformation of SHP2, the N-SH2 domain interacts with the catalytic PTP domain and blocks access to the active site, while the interface between the N-SH2, C-SH2, and PTP domains forms the canonical SHP099 allosteric tunnel \cite{chen2016shp2}. Although SHP099 did not recover the reference pose, the four-state ligand-contact model resolved macrostates overlapping two experimentally established allosteric regions. M2 occupies the canonical SHP099 tunnel near His114 and Thr253 but remains displaced and differently oriented relative to the reference pose, whereas M0 overlaps the distinct N-SH2--PTP latch site near Tyr80 and Lys280, which is targeted by the allosteric inhibitors SHP244 and SHP504 \cite{fodor2018dual}. M1 and M3 instead represent exposed PTP and C-SH2 surface contacts near Leu436 and Tyr197/Glu204, respectively. SHP099 therefore samples established allosteric regions without reproducing its reference binding geometry. Under the applied C$\alpha$ restraints, the SHP2 domains showed limited internal deformation and relative motion, while pocket RMSDs and $Q_{\mathrm{pocket}}$ remained approximately stationary after an initial relaxation (Supplementary Figure~\ref{sup_fig:shp2_rearrangement}).

Ligand 43 belongs to a series of type-I inhibitors that bind the CDK8 ATP site \cite{schiemann2016cdk8}. None of the 25 trajectories recovered the reference pose, and the three-state ligand-contact model did not resolve an ATP-pocket macrostate. M0 represents a distal CDK8 surface contact near Pro124, M1 samples a region adjacent to CDK8 Trp146 and Cyclin C Arg58, and M2 lies on the Cyclin C surface near Asp182 (Figure~\ref{fig:additional_binding_systems}f). These states therefore describe recurrent off-site contacts with limited specificity for both the target chain and the reference pocket.

\section{Conclusion}

\MODELNAME{} extends the \textit{noh} framework from protein conformational dynamics to a shared heavy-atom coarse-grained potential for chemically diverse biomolecular systems. By retaining heavy atoms while encoding local hydrogenation and total system charge, the representation provides a common chemical description across these molecular classes without separate component potentials or fixed energetic priors. Within this representation, a single learned potential supports folded protein structures and conformational landscapes, stable small-molecule dynamics, and simulations of both binary protein--ligand and ternary protein--protein--ligand complexes initiated from bound and unbound ligand configurations.


The protein--ligand benchmark shows that \MODELNAME{} preserves many preformed complexes, although in HIF2A the internally structured partner chain reoriented relative to the ligand-binding monomer, causing substantial loss of the reference interface contacts. The unbound binding trials provide a stricter test of interaction specificity: \MODELNAME{} recovered the bound pose from bulk for trypsin--benzamidine and TYK2, sampled established binding regions without reproducing the SHP099 pose in SHP2, and predominantly formed off-site contacts in CDK8.

Within the trypsin--benzamidine case study, the contrasting behavior of benzamidinium and neutral benzamidine further supports the chemical specificity of the learned potential. Greater bound-pose stability of benzamidinium is accompanied by more persistent Asp171 contacts, whereas neutral benzamidine shows reduced contact occupancy even without clear unbinding (Figure~\ref{fig:benzamidine_rmsd}a,c).
Beyond protonation sensitivity, the trypsin--benzamidine simulations also demonstrate improved binding-sampling efficiency: at a fixed estimated GPU-time budget of 4~hours per trajectory, \MODELNAME{} recovered the bound pose in 3.3-fold more trajectories than all-atom MM (Figure~\ref{fig:tryps_binding_msm}a). Thus, the learned coarse-grained potential improves bound-pose sampling per unit of computational time despite its lower simulated-time throughput.

These simulations also support MSM-based estimation of protein--ligand binding free energies with the shared potential. For the calcium-containing 3PTB system, \MODELNAME{} places the bound and bulk regions in similar locations to the all-atom model, but the bulk region lies higher in free energy relative to the bound basin (Figure~\ref{fig:tryps_binding_msm}b). This is consistent with the more favorable binding free-energy estimate and suggests that \MODELNAME{} may over-stabilize the bound state relative to bulk. The \MODELNAME{} binding free-energy estimates are more favorable than the experimental references for both trypsin--benzamidine and TYK2--EJM31. The trypsin estimates vary by 0.06~kcal/mol or less over bulk cutoffs of 8--20~\AA{}, whereas the TYK2 estimate shifts overall toward more negative values over 10--15~\AA{} (Supplementary Figure~\ref{sup_fig:dg_vs_bulk_cutoff}).
%
For CDK8, ligand 43 predominantly samples off-site configurations on both CDK8 and Cyclin C rather than concentrating around the ATP-binding region expected for this type-I inhibitor \cite{schiemann2016cdk8}. Two recurrent configurations occur near Cyclin C Arg58 and Asp182, residues belonging to a conserved Cyclin C-specific surface groove proposed to mediate protein interactions \cite{hoeppner2005structure}. These contacts are therefore consistent with off-site encounters at a chemically plausible surface region. However, persistent off-site contacts suggest that some protein-surface interactions may be over-stabilized relative to pocket binding. 

Overall, \MODELNAME{} combines a chemically informative representation with total-charge conditioning in a compact architecture. Together with the optimized implementation, this framework makes long-timescale biomolecular simulations practical for large systems while retaining chemical specificity.
\section{Methods}

\subsection{Training procedure}

We trained \MODELNAME{} by force matching \cite{noid2008multiscale,izvekov2005multiscale} to the aggregated \textit{noh} forces using TorchMD-Net \cite{pelaez2024torchmd}. The objective comprised a Huber loss with $\delta=5$~kcal/mol/\AA{} on the Cartesian force errors and no energy term. To reduce the contribution of larger systems, we weighted each molecular graph's loss by $1/\sqrt{N}$, where $N$ is the number of \textit{noh} sites.

Training proceeded sequentially on small molecules, protein conformations from mdCATH \cite{mirarchi2024mdcath}, and protein--ligand complexes. Reference forces were calculated from quantum-mechanical calculations on isolated small molecules, whereas protein and protein--ligand
force labels were obtained from molecular-mechanics force fields for configurations sampled in solvated molecular dynamics. The large differences in system size caused substantial variation in memory use for batches containing a fixed number of molecular graphs; batches containing several protein--ligand complexes could exceed the 24~GB memory available on each NVIDIA RTX 4090 GPU. Separate training stages therefore used batch sizes of 128, 12, and 2 for small molecules, proteins, and protein--ligand complexes, respectively. We denote the small-molecule checkpoint as \MODELNAME{} (SM) and the final checkpoint as \MODELNAME{}. We describe dataset construction, subset selection, and hyperparameters in the Supplementary Information and Supplementary Table~\ref{sup_tab:model_hparams}.

\subsection{Molecular dynamics simulations}

We ran \MODELNAME{} simulations in OpenMM 8.5 \cite{eastman2023openmm} using the Acellera-TensorNet plugin, available through PlayMolecule AI (\url{https://playmolecule.ai}). Unless otherwise stated, systems were minimized with L-BFGS \cite{liu1989limited} for at most 2,000 iterations and simulated in the NVT ensemble at 300~K using a Langevin middle integrator with a friction coefficient of 1~ps$^{-1}$, a 4~fs time step, and coordinates saved every 0.1~ns. Simulations were non-periodic, and no explicit long-range electrostatic term was considered.

\noindent The $\alpha$3D and villin simulations followed the AMARO v1.0 comparison analysis \cite{mirarchi2024amaro} and were propagated with TorchMD at 350~K for 8 and 10~ns, respectively \cite{doerr2020torchmd}.

For the binding-reconstruction and binding-exploration simulations initiated from unbound ligand configurations, we applied positional harmonic restraints to protein C$\alpha$ atoms more than 9~\AA{} from the reference ligand pose, with $k_{\mathrm{CA}}=0.5$~kcal\,mol$^{-1}$\,\AA{}$^{-2}$.
A flat-bottom potential restricted ligand diffusion to a defined region around the protein:

\begin{equation}
U(\Delta \mathbf{r}) =
\frac{k_{\mathrm{FB}}}{2}
\sum_{i \in \{x,y,z\}}
\left[
\max\left(|\Delta r_i|-h_i,0\right)
\right]^2 ,
\end{equation}

where $\Delta\mathbf{r}=\mathbf{R}_{L}-\mathbf{R}_{P}$, $k_{\mathrm{FB}}=0.1$~kcal\,mol$^{-1}$\,\AA{}$^{-2}$, and $h_i$ is the half-width of the allowed region along axis $i$. For all systems, $\mathbf{R}_{L}$ denotes the ligand heavy-atom center of mass and $\mathbf{R}_{P}$ the centroid of the selected pocket C$\alpha$ atoms. Supplementary Table~\ref{sup_tab:binding_restraints} reports the restraint dimensions and pocket definitions.

\subsection{Small-molecule benchmarks}
The AceFF v1.0 and v2.0 checkpoints were obtained from Hugging Face (\href{https://huggingface.co/Acellera/AceFF-1.0}{v1.0}; \href{https://huggingface.co/Acellera/AceFF-2.0}{v2.0}). For the dynamics benchmark, we ran AceFF v2.0 without constraints at 1~fs, with constraints on bonds involving hydrogen at 2~fs, and with the same constraints combined with hydrogen mass repartitioning at 4~fs \cite{hopkins2015long}. No bond constraints or mass repartitioning were used for the \textit{noh} models. We compared heavy-atom local-geometry distributions with matched AceFF v2.0 simulations using the first Wasserstein distance, $W_{1}$, with a leave-one-replica-out AceFF baseline; more details are provided in the Supplementary Information.

\subsection{Protein interface characterization}
We defined a native interchain residue contact when any pair of heavy atoms from two different chains was within 5~\AA{} in the reference structure. Interface retention was quantified as

\begin{equation}
Q_{\mathrm{interface}}(t) = \frac{N_{\mathrm{retained}}(t)}
     {N_{\mathrm{native}}},
\end{equation}

where $N_{\mathrm{native}}$ is the number of native interchain residue contacts and $N_{\mathrm{retained}}(t)$ is the number of those contacts retained at time $t$.
For HIF2A, we separated partner-chain conformational changes from relative interchain motion by comparing its C$\alpha$-RMSD after self-alignment and after alignment on the ligand-binding monomer. We quantified relative interchain reorientation from the rigid-body rotation of the partner core after aligning it to the structured core of the ligand-binding chain.

\subsection{Binding analyses}

The prepared 1BTY and 3PTB complexes both had a total system charge of $+9$ with benzamidinium ($+1$), or $+8$ with neutral benzamidine. The 1BTY system contained no calcium ion. Relative to 1BTY, neutral His57 in 3PTB lowered the protein charge by one unit, and deprotonation of the corresponding glutamate (neutral, protonated GLH52 in 1BTY; GLU70 in 3PTB) lowered it by another. The retained Ca$^{2+}$ ion in 3PTB balanced this two-unit reduction in protein charge.

For the trypsin--benzamidine binding-reconstruction simulations, unbound starting configurations were selected without conditioning on whether the corresponding all-atom trajectory subsequently reached the bound state. For the 50~ns comparison, trajectories without pose recovery were right-censored at 50~ns in both ensembles. Pointwise 95\% confidence intervals for cumulative pose-recovery fractions were calculated using Wilson score intervals \cite{wilson1927probable}. Confidence intervals for differences between models at 50~ns and 4~GPU hours were obtained from 100,000 paired bootstrap resamples of the 100 matched trajectory pairs, using the 2.5th and 97.5th percentiles.

To evaluate trypsin--benzamidine pose recovery at a fixed GPU-time budget per trajectory (Figure~\ref{fig:tryps_binding_msm}a), we converted simulation time to estimated GPU time as $t_{\mathrm{GPU}}\,[\mathrm{h}]=24\,t_{\mathrm{sim}}\,[\mathrm{ns}]/v$, using median throughputs $v$ of 592~ns/day for all-atom MM and 164~ns/day for \MODELNAME{}, both measured on an NVIDIA RTX 4090 GPU.
At each GPU-time budget, we calculated cumulative pose-recovery fractions from first-passage events within the corresponding model-specific simulation windows, including all-atom observations beyond the 50~ns endpoint used for the fixed-length analysis.

\subsection{Markov state models}

We projected the input features with time-lagged independent component analysis (TICA) \cite{perezh2013tica}, discretized the projected trajectories using mini-batch $k$-means clustering, estimated reversible Markov state models (MSMs) \cite{chodera2014markov}, and coarse-grained the resulting microstates with PCCA+ \cite{deuflhard2005pcca}. Supplementary Table~\ref {sup_tab:msm_parameters} reports system-specific feature definitions, TICA and MSM lag times, retained TICA components, and the numbers of microstates and macrostates.

For $\alpha$3D and villin, we followed the landscape-analysis workflow used for AMARO v1.0 \cite{mirarchi2024amaro}, discarding the first 10\% of each trajectory and analyzing the remaining frames with a stride of 5. TICA was fitted to pairwise C$\alpha$ distances from the all-atom reference ensemble \cite{majewski2023machine}, and \MODELNAME{} trajectories were projected onto the same basis.

For the ligand-binding models, we used continuous distances between the ligand center of mass and the center of mass of each protein residue, capped at 25~\AA{}. For trypsin--benzamidine, TICA was fitted to the 491 all-atom trajectories and the \MODELNAME{} binding-reconstruction trajectories were projected onto the same basis. The two ensembles were then discretized independently. For TYK2, SHP2, and CDK8, we performed TICA and clustering independently for each system, distinguishing CDK8 residues by chain.
Ligand--protein contacts were evaluated after MSM construction using heavy-atom cutoffs of 4.5~\AA{} for TYK2 and SHP2 and 4.0~\AA{} for CDK8. CDK8 ATP-pocket contacts were evaluated against residues located within 5~\AA{} of the ligand in the reference structure.

For binding free-energy calculations, a frame was classified as bulk-like when its ligand COM was farther than the selected cutoff from every protein-residue COM. For trypsin--benzamidine, we split each mixed microstate into bulk and nonbulk components before fitting the MSM and PCCA+ partition. We then grouped all wholly bulk microstates into a dedicated bulk macrostate, leaving the fitted microstate transition matrix unchanged (Supplementary Figure~\ref{sup_fig:tryps_macrostate_composition}). This macrostate served as the kinetic source; the nonbulk macrostate with the highest fraction of reference-pose frames (protein-aligned ligand RMSD $\leq2.5$~\AA{}) served as the bound sink. We tested bulk cutoffs of 4.5--20~\AA{} and used 12~\AA{} for the reported estimates. Binding free energies were calculated at 300~K as

\begin{equation}
\Delta G^\circ = -RT
\ln\left[
    \frac{p_{\mathrm{bound}}}{p_{\mathrm{bulk}}}
    \frac{V_{\mathrm{bulk,eff}}}{V^\circ}
\right],
\end{equation}

where $p_{\mathrm{bound}}$ and $p_{\mathrm{bulk}}$ are the stationary probabilities of the complete assigned macrostates, $R$ is the gas constant, and $V^\circ=1660.54$~\AA{}$^3$ is the molecular volume corresponding to the 1~M standard state. We evaluated the cutoff-specific effective bulk volume, $V_{\mathrm{bulk,eff}}$, over the region satisfying the bulk criterion. State populations and sampling diagnostics are reported in Supplementary Table~\ref{sup_tab:tyk2_equilibrium_diagnostics}.

Uncertainties were estimated by resampling complete trajectories with replacement and repeating the analysis for 100 bootstrap resamples per trypsin model and 20 for TYK2. Reported uncertainties are sample standard deviations of finite bootstrap estimates.

\section{Data availability}
The mdCATH dataset is publicly available \cite{mirarchi2024mdcath}, whereas the in-house small-molecule and protein--ligand training datasets are not publicly released. The final \MODELNAME{} checkpoint is available in the \href{https://huggingface.co/compsciencelab/AMARO2}{AMARO2 repository on Hugging Face}.

\section{Code availability}
The TorchMD-Net implementation of \MODELNAME{}, together with scripts for system preparation and OpenMM simulations using PythonForce, is available at \url{https://github.com/compsciencelab/amaro}. The faster Acellera-TensorNet implementation is accessible through the \href{https://playmolecule.ai}{PlayMolecule AI interface}.

\bibliographystyle{naturemag}
\bibliography{reference}

\section{Acknowledgements}
A.M. is financially supported by the Generalitat de Catalunya's Agency for Management of University and Research Grants (AGAUR) PhD grant 2026 FI-3-00278.

\section{Author contributions}
A.M. and G.D.F. designed the method and wrote the manuscript. A.M., S.E.F., and A.P. developed the datasets and computational tools. A.M. and A.P. conducted the analyses. G.D.F. supervised the work.

\section{Competing interests}
The authors declare the following competing financial interests: S.E.F., A.P., and
G.D.F. have a potential conflict of interest due to direct interests in Acellera (see affiliations).

\onecolumn
\renewcommand{\theequation}{S\arabic{equation}}
\renewcommand{\thetable}{S\arabic{table}}
\renewcommand{\thefigure}{S\arabic{figure}}

\setcounter{equation}{0}
\setcounter{table}{0}
\setcounter{figure}{0}

\section*{Supplementary Information}

\subsection*{\MODELNAME{} hyperparameters}
Table~\ref{sup_tab:model_hparams} compares the \MODELNAME{} architecture and training settings with AMARO v1.0. We applied a Huber loss of $\delta=5$~kcal/mol/\AA{} to mapped Cartesian force errors and used no energy loss. To limit the influence of larger graphs, we weighted each loss by $1/\sqrt{N}$, where $N$ is the number of beads. We also replaced the cosine cutoff \cite{behler2007generalized} with a polynomial cutoff and used two message-passing layers with a 6~\AA{} cutoff. Reducing the embedding dimension and radial basis size gives 236,000 parameters, compared with 576,000 in AMARO v1.0. At each training stage, we selected the checkpoint with the lowest total validation Huber loss and applied a learning-rate warm-up when training resumed to stabilize the initial optimization stage.

\begin{table}[!ht]
    \centering
    \small
    \caption{Comparison of AMARO v1.0 and \MODELNAME{} hyperparameters. \MODELNAME{} was trained sequentially on small molecules, mdCATH proteins, and protein--ligand systems using four NVIDIA RTX 4090 GPUs at every stage.}
    \label{sup_tab:model_hparams}
    \begin{tabular}{p{0.32\linewidth}p{0.24\linewidth}p{0.34\linewidth}}
    \hline
    \textbf{Hyperparameter} & \textbf{AMARO v1.0} & \textbf{\MODELNAME{}} \\
    \hline
    \multicolumn{3}{l}{\textbf{Neural network architecture}} \\
    activation & SiLU & SiLU \\
    aggr & add & add \\
    cutoff\_lower & 0.0 \AA & 0.0 \AA \\
    cutoff\_upper & 5.0 \AA & 6.0 \AA \\
    embedding\_dimension & 128 & 64 \\
    equivariance\_invariance\_group & O(3) & O(3) \\
    max\_num\_neighbors & 64 & 96 \\
    max\_z & 100 & 32 \\
    num\_layers & 1 & 2 \\
    num\_rbf & 32 & 24 \\
    precision & float32 & float32 \\
    q\_dim & --- & 16 \\
    rbf\_type & expnorm & gauss \\
    \hline
    \multicolumn{3}{l}{\textbf{Training}} \\
    batch size & 8 & 128, 12, and 2 for the small-molecule, protein, and protein--ligand stages \\
    distance\_influence & both & both \\
    early\_stopping\_patience & 30 & 30 \\
    ema\_alpha\_neg\_dy & 1.0 & 1.0 \\
    ema\_alpha\_y & 1.0 & 1.0 \\
    learning rate & 0.0003 & 0.001, 0.0005, and 0.0003 for the small-molecule, protein, and protein--ligand stages \\
    max\_num\_epochs & 100 & 1000 \\
    neg\_dy\_weight & 1.0 & 1.0 \\
    test\_interval & -1 & -1 \\
    test\_size & 0.1 & null \\
    train\_size & null & 0.8 \\
    trainable\_rbf & false & false \\
    val\_size & 0.05 & 0.1 \\
    weight\_decay & null & 0.0005 \\
    y\_weight & 0.0 & 0.0 \\
    \hline
    \end{tabular}
\end{table}

\FloatBarrier

\subsection*{Training datasets}

\subsubsection*{Small molecules}
We used the AceFF dataset \cite{farr2026aceff}, which includes molecules from PubChem \cite{kim2025pubchem}. The in-house dataset contains approximately 2 million unique molecules and 12 million structures. Molecules contain up to 30 atoms, carry total charges from $-2$ to $+2$, and include H, B, C, N, O, F, Si, P, S, Cl, Br, and I. Electronic energies and atomic forces were computed at the $\omega$B97M-V/def2-TZVPPD level \cite{mardirossian2016wb97mv} using PySCF \cite{sun2020pyscf} and GPU4PySCF \cite{li2024gpu4pyscf,wu2024gpu4pyscf}; only forces were retained for training. The structures include minimized and high-energy geometries. We used this in-house dataset because it retains the connectivity and SMILES information required for the \textit{noh} mapping. We excluded conformers with any atomic force norm above 700~kcal/mol/\AA{}, leaving approximately 10 million conformers for the small-molecule training stage.

\subsubsection*{Proteins}
We selected 2,659 of the 5,398 mdCATH proteins~\cite{mirarchi2024mdcath} by retaining systems with fewer than 250 residues and included higher-temperature conformations to broaden structural coverage~\cite{mirarchi2024amaro}. We generated the 300~K data using the molecular dynamics setup described in the original mdCATH study. After excluding conformations with any atomic force norm above 300~kcal/mol/\AA{}, the resulting training dataset comprised 1,010,318 conformations at 300~K, 542,325 at 320~K, 180,866 at 348~K, and 90,959 at 379~K.

\subsubsection*{Protein--ligand complexes}
Protein--ligand conformations were taken from Acellera \textit{HoloSet}, an internal dataset which comprises simulations of 3,758 unique protein--ligand complexes at 310~K. Simulations were performed with ACEMD \cite{harvey2009acemd} using a Langevin damping coefficient of 0.1~ps$^{-1}$, particle-mesh Ewald electrostatics with a 9~\AA{} real-space cutoff, and van der Waals switching between 7.5 and 9~\AA{}. Each system was simulated for 100~ns with a 4~fs time step, and coordinates were saved every 100~ps. We then recomputed forces for the saved configurations using OpenMM \cite{eastman2023openmm}. Proteins were parameterized with Amber ff19SB \cite{tian2020ff19sb}, and ligands with GAFF2 \cite{wang2004gaff,ambergaff2} and AM1-BCC partial charges \cite{jakalian2000am1bcc,jakalian2002am1bcc}. Complexes were solvated with TIP3P water \cite{jorgensen1983comparison}. Conformations containing any atomic force norm above 300~kcal/mol/\AA{} were excluded, leaving approximately 3.5 million conformations, from which 500,000 were selected for training using a stride of 7. The 1BTY trypsin--benzamidine complex was present in the source dataset but excluded from training.

\FloatBarrier

\subsection*{Small-molecule validation}

\subsubsection*{Simulation stability}

\begin{figure}[ht]
    \centering
    \includegraphics[width=0.85\linewidth]{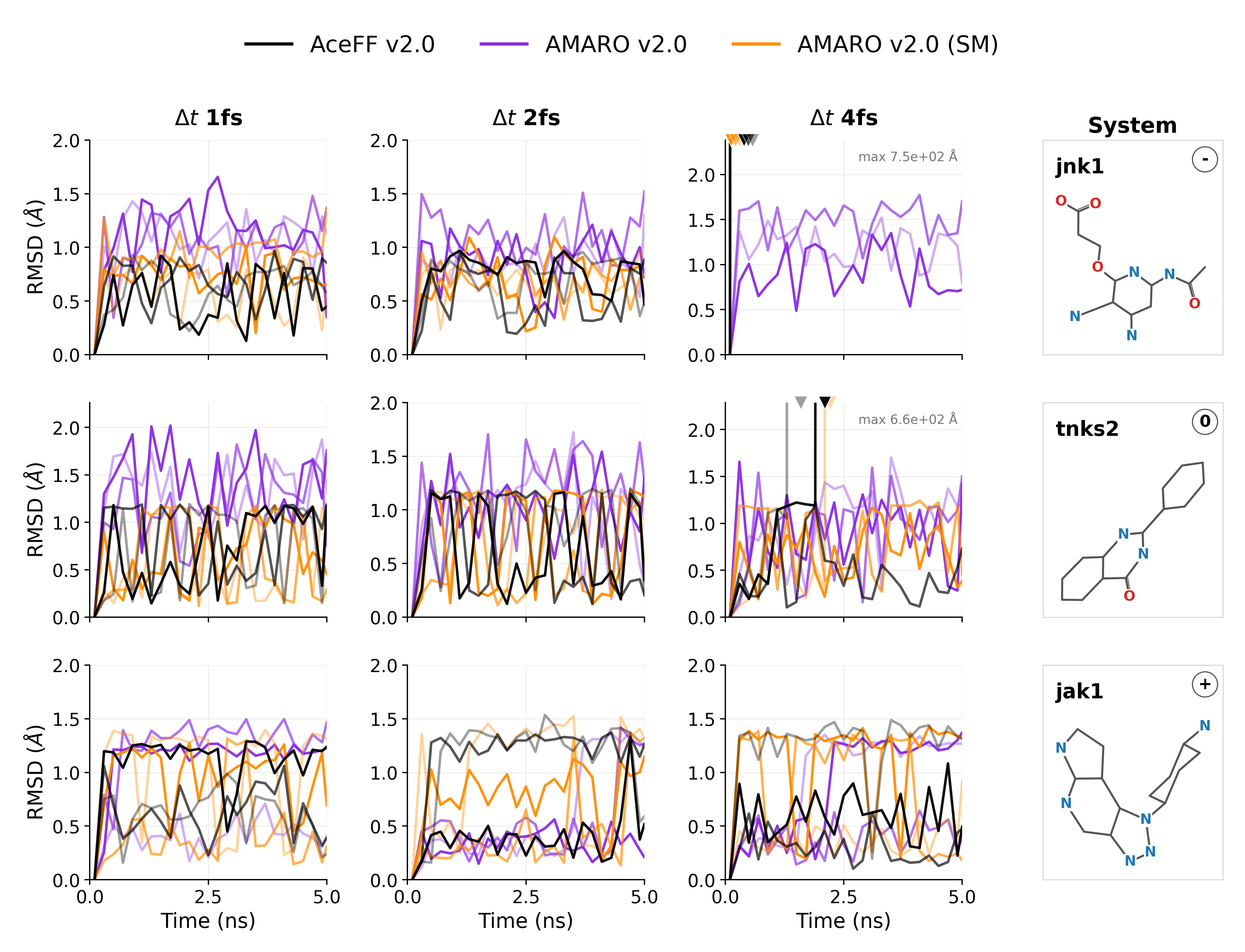}
    \caption{Heavy-atom RMSD relative to the initial conformer during 5~ns simulations of ligands from the JNK1 ($-1$), TNKS2 (0), and JAK1 ($+1$) benchmark sets; parentheses indicate formal charges. Three independent replicas were run for AceFF v2.0, \MODELNAME{} (SM), and \MODELNAME{} at 300~K with time steps of 1, 2, and 4~fs. AceFF v2.0 used no constraints at 1~fs, constraints on bonds involving hydrogen at 2~fs, and these constraints combined with hydrogen mass repartitioning at 4~fs. We used neither bond constraints nor mass repartitioning for the \textit{noh} models. Downward markers indicate divergent trajectories; gray labels report RMSD maxima above the plotted range.}
    \label{sup_fig:dynamic_small_mol}
\end{figure}

\subsubsection*{Local-geometry analysis}

We compared heavy-atom local-geometry distributions with matched AceFF v2.0 simulations using a common heavy-atom connectivity graph derived from the source SDF. Bond angles were defined by pairs of neighbors around each atom, and out-of-plane angles were evaluated at three-connected centers carrying a multiple or aromatic bond. The 1--4 distances were measured between atoms separated by exactly three bonds.

We analyzed complete trajectories with a maximum heavy-atom RMSD below 10~\AA{}, using frames from 0.5 to 5~ns. For each molecule and time step, we compared each replica of the two \textit{noh} models with the pooled AceFF replicas. Reference variability was assessed by comparing each AceFF replica with the remaining reference replicas (leave-one-replica-out). Conditions with fewer than two usable AceFF replicas were excluded.

Distribution agreement was quantified using the first Wasserstein distance,
$W_1(P,Q)=\int |F_P(x)-F_Q(x)|\,dx$,
where $F_P$ and $F_Q$ are the cumulative distribution functions. This metric captures differences in distribution location and width, and retains the coordinate units (\AA{} for distances and degrees for angles).

\begin{figure}[ht]
    \centering
    \includegraphics[width=0.98\linewidth]{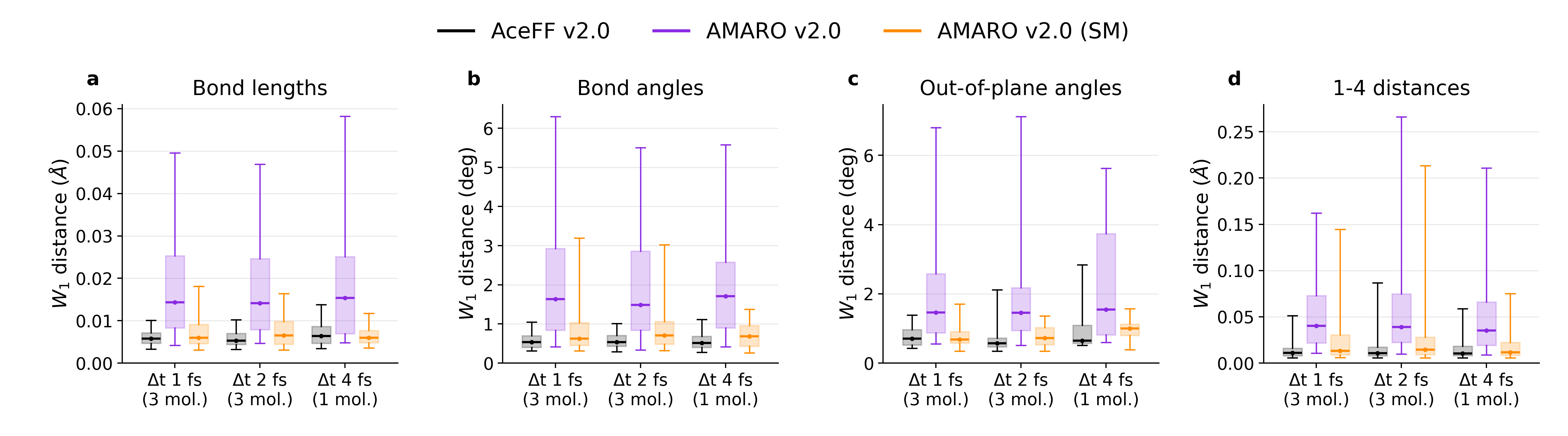}
    \caption{Heavy-atom local-geometry comparison with AceFF v2.0. Wasserstein-1 ($W_1$) distances are shown for (a) bond lengths, (b) bond angles, (c) out-of-plane angles at nominally planar centers, and (d) 1--4 distances. Black, purple, and orange denote AceFF v2.0, \MODELNAME{}, and \MODELNAME{} (SM), respectively. Boxes show the interquartile range, dots the medians, and whiskers the 5th--95th percentiles across coordinate--replica comparisons. AceFF values represent the leave-one-replica-out baseline. Counts beneath each time step indicate the contributing molecules; only JAK1 contributes at 4~fs because the other conditions had fewer than two usable AceFF replicas. Analysis used the 0.5--5~ns window and trajectory-selection criteria described in this section.}
    \label{sup_fig:small_mol_geometry}
\end{figure}

\FloatBarrier



\FloatBarrier

\subsection*{TICA and MSM parameters}

\begin{table}[!htbp]
    \centering
    \scriptsize
    \caption{System-specific TICA and MSM parameters. AA denotes the all-atom reference, and COM denotes the center of mass. For the ligand-binding models, microstate counts refer to the original clustering, before any bulk/nonbulk splitting. Trypsin macrostate counts refer to the final partition: both models requested two PCCA+ states and contained three final macrostates after grouping wholly bulk microstates into a dedicated bulk state. For the other ligand-binding systems, macrostate counts refer to the original PCCA+ partition.}
    \label{sup_tab:msm_parameters}
    \resizebox{0.95\textwidth}{!}{%
    \begin{tabular}{@{}llllrrr@{}}
        \toprule
        System & Features & TICA lag (ns) & TICs & Microstates & MSM lag (ns) & Macrostates \\
        \midrule
        \multicolumn{7}{l}{\textbf{Protein conformational landscapes}} \\
        $\alpha$3D & pairwise C$\alpha$ distances & 2 & 4 (AA), 3 (\MODELNAME{}) & 600 & 2 & 3 \\
        Villin & pairwise C$\alpha$ distances & 2 & 4 (AA), 3 (\MODELNAME{}) & 250 & 1 & 3 \\
        \addlinespace
        \multicolumn{7}{l}{\textbf{Ligand-contact models}} \\
        Trypsin AA & ligand--residue COM distances & 2 & 3 & 400 & 20 & 3 \\
        Trypsin \MODELNAME{} & ligand--residue COM distances & 2 & 3 & 200 & 10 & 3 \\
        TYK2 & ligand--residue COM distances & 2 & 3 & 100 & 20 & 5 \\
        SHP2 & ligand--residue COM distances & 2 & 3 & 200 & 10 & 4 \\
        CDK8 & ligand--residue COM distances & 1 & 3 & 100 & 20 & 3 \\
        \bottomrule
    \end{tabular}%
    }
\end{table}

\begin{table}[!htbp]
\centering
\caption{State and sampling diagnostics for the 12~\AA{} bulk cutoff. Pose purity is the fraction of frames in the selected bound macrostate with protein-aligned ligand RMSD $\leq2.5$~\AA{}. Lagged pairs count bulk-to-bound and bound-to-bulk transitions between frames separated by the selected MSM lag.}
\label{sup_tab:tyk2_equilibrium_diagnostics}
\begin{tabular}{@{}lrrr@{}}
\toprule
Diagnostic & Trypsin AA & Trypsin \MODELNAME{} & TYK2 \MODELNAME{}\\
\midrule
MSM lag / clusters & 20 ns / 400 & 10 ns / 200 & 20 ns / 100\\
Active-frame coverage & 99.9994\% & 99.9994\% & 98.3168\%\\
$p_{\rm bulk}$ & 0.01253 & $1.105\times10^{-6}$ & $9.07\times10^{-8}$\\
$p_{\rm bound}$ & 0.97968 & 0.99806 & 0.02967\\
Pose purity & 91.23\% & 93.78\% & 38.15\%\\
$V_{\rm bulk,eff}$ (\AA{}$^3$) & $381296\pm155$ & $510063\pm340$ & $481728\pm266$\\
Lagged bulk$\rightarrow$bound / bound$\rightarrow$bulk & 5896 / 302 & 1416 / 0 & 333 / 5\\
\bottomrule
\end{tabular}
\end{table}

\begin{figure}[!htbp]
    \centering
    \begin{subfigure}[b]{0.31\textwidth}
        \centering
        \includegraphics[width=\linewidth]{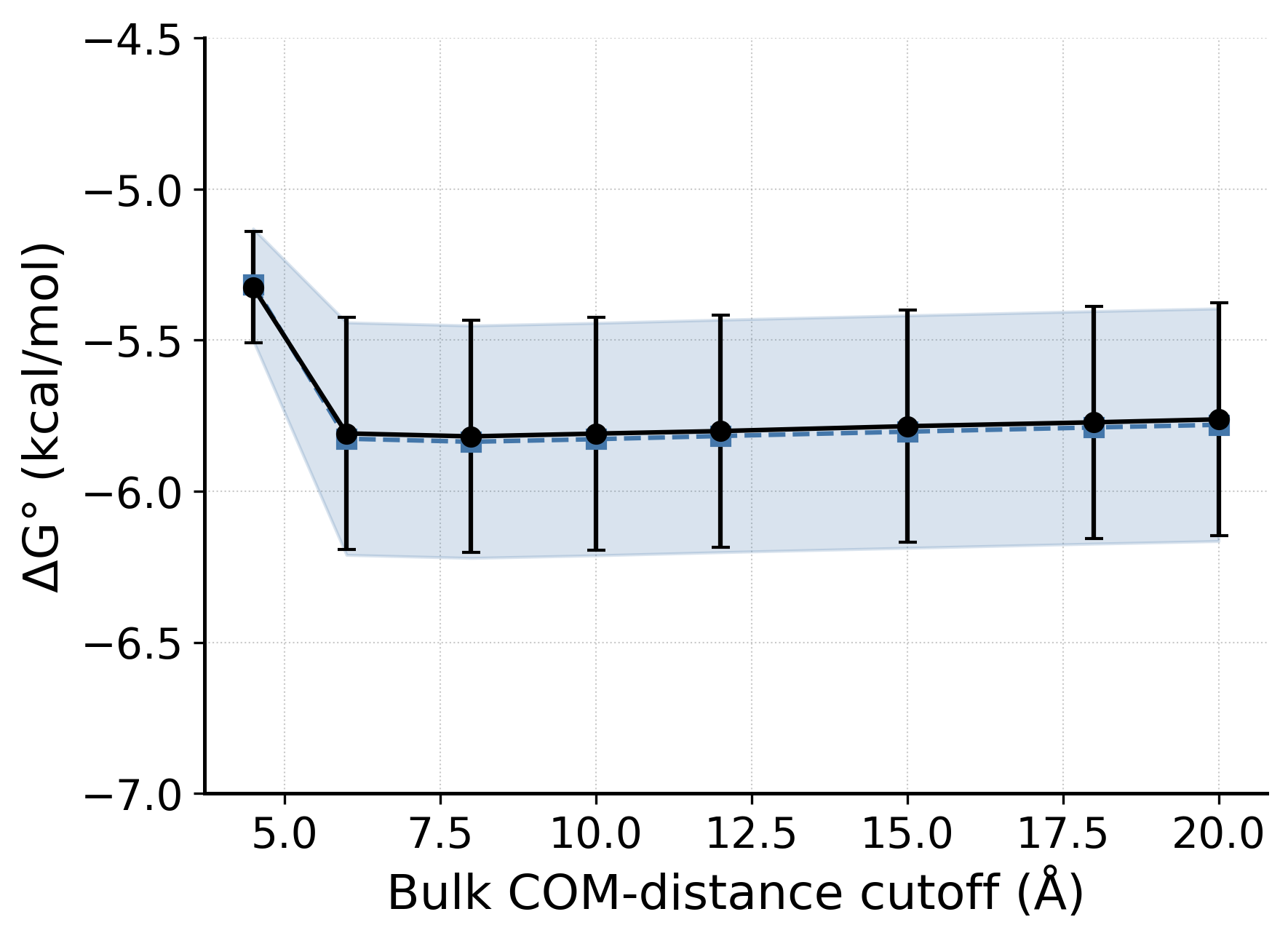}
        \caption{3PTB all-atom}
    \end{subfigure}
    \hfill
    \begin{subfigure}[b]{0.31\textwidth}
        \centering
        \includegraphics[width=\linewidth]{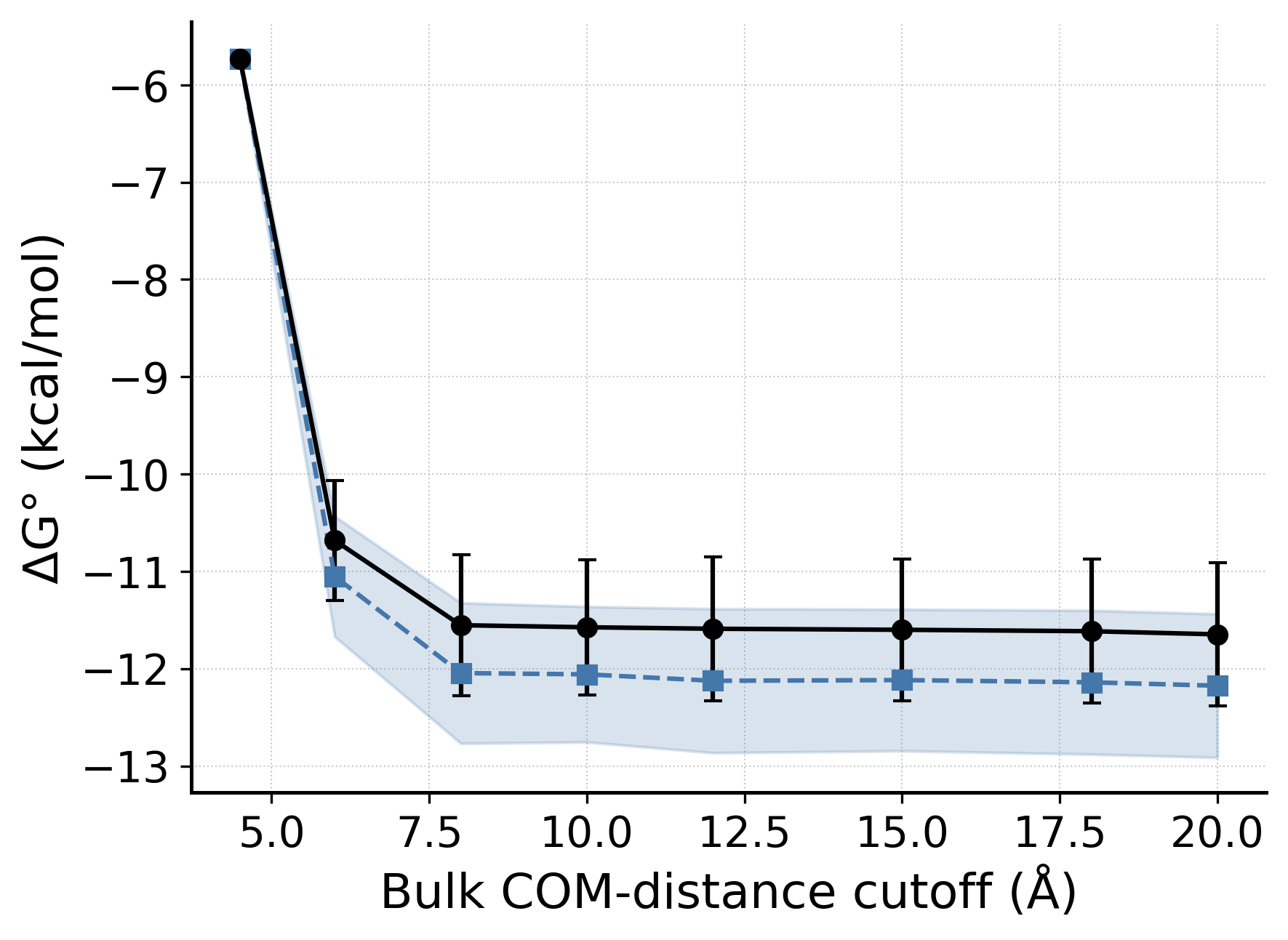}
        \caption{3PTB \MODELNAME{}}
    \end{subfigure}
    \hfill
    \begin{subfigure}[b]{0.31\textwidth}
        \centering
        \includegraphics[width=\linewidth]{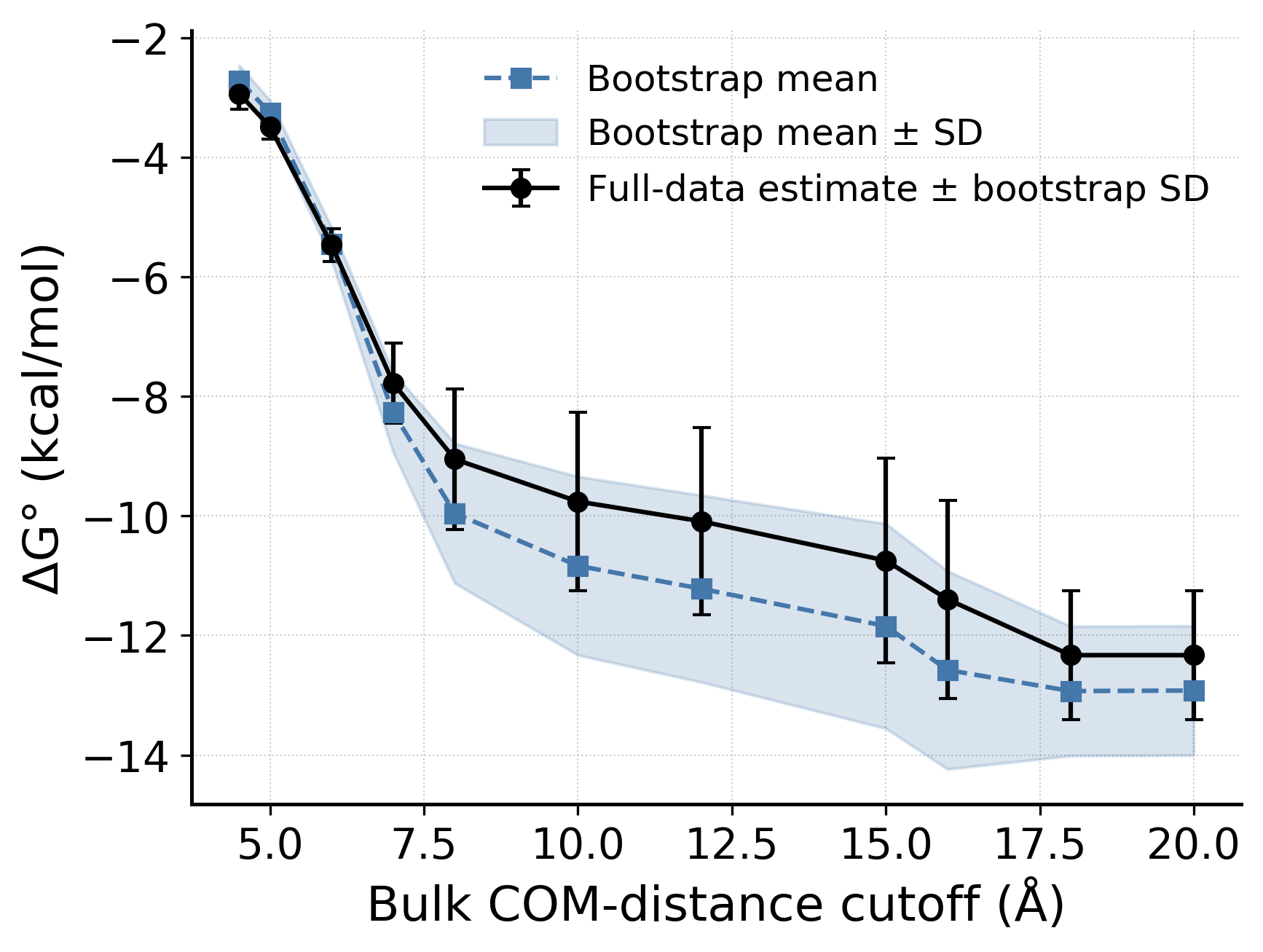}
        \caption{TYK2 \MODELNAME{}}
    \end{subfigure}
    \caption{Binding free energy as a function of the operational bulk-state cutoff for (a) all-atom trypsin--benzamidine, (b) \MODELNAME{} trypsin--benzamidine, and (c) \MODELNAME{} TYK2--EJM31. We classified a frame as bulk-like when its ligand COM was farther than the indicated cutoff from every protein-residue COM; equivalently, the minimum ligand-COM--residue-COM distance exceeded the cutoff. For trypsin, mixed microstates were split into bulk and nonbulk microstates before fitting the MSM and PCCA+ partition; wholly bulk microstates were then grouped into the dedicated bulk-source macrostate without changing the fitted microstate transition matrix. The plotted free energies use the cutoff-specific effective bulk volume. The manuscript values use a common 12~\AA{} cutoff; the TYK2 20~ns/100-cluster model is supported over 10--15~\AA{}, while larger cutoffs have insufficient connectivity. Black circles show estimates from all trajectories; dashed blue squares show the mean of finite bootstrap refits, the blue band shows the bootstrap mean $\pm$ sample SD, and black error bars show the full-data estimate $\pm$ bootstrap SD. The trypsin models used 100 refits per cutoff, whereas TYK2 used 20 refits. Failed estimates were excluded rather than assigned zero.}
    \label{sup_fig:dg_vs_bulk_cutoff}
\end{figure}
\FloatBarrier

\subsection*{Binding-simulation restraints}
\begin{table}[!htbp]
    \centering
    \small
    \caption{Flat-bottom restraint definitions for the binding simulations. Full dimensions are reported in \AA{}, and each half-width $h_i$ is one-half of the corresponding dimension. The trypsin--benzamidine dimensions are means across the 100 starting configurations; the other systems used fixed cubic dimensions.}
    \label{sup_tab:binding_restraints}
    \resizebox{0.95\textwidth}{!}{%
    \begin{tabular}{@{}lll@{}}
        \toprule
        System & Full dimensions (\AA{}) & Pocket-centroid C$\alpha$ residues\\
        \midrule
        Trypsin--benzamidine & $77.1\times77.5\times83.7$ & Asp171, Ser172, Trp193, Gly194, Ser195, Gly196 \\
        TYK2--EJM31 & $80.0\times80.0\times80.0$ & Glu905, Ala928, Val981, Pro982, Asn1028 \\
        SHP2--SHP099 & $146.4\times146.4\times146.4$ & Glu110, Arg111, Phe113, Glu250, Thr253 \\
        CDK8--ligand 43 & $155.0\times155.0\times155.0$ & CDK8 chain A: Tyr32, Asp98, Ala100, Asn156, Asp173 \\
        \bottomrule
    \end{tabular}%
    }
\end{table}

\FloatBarrier

\subsection*{Trypsin--benzamidine pose recovery}
\begin{figure}[!htbp]
    \centering
    \includegraphics[width=0.6\textwidth]{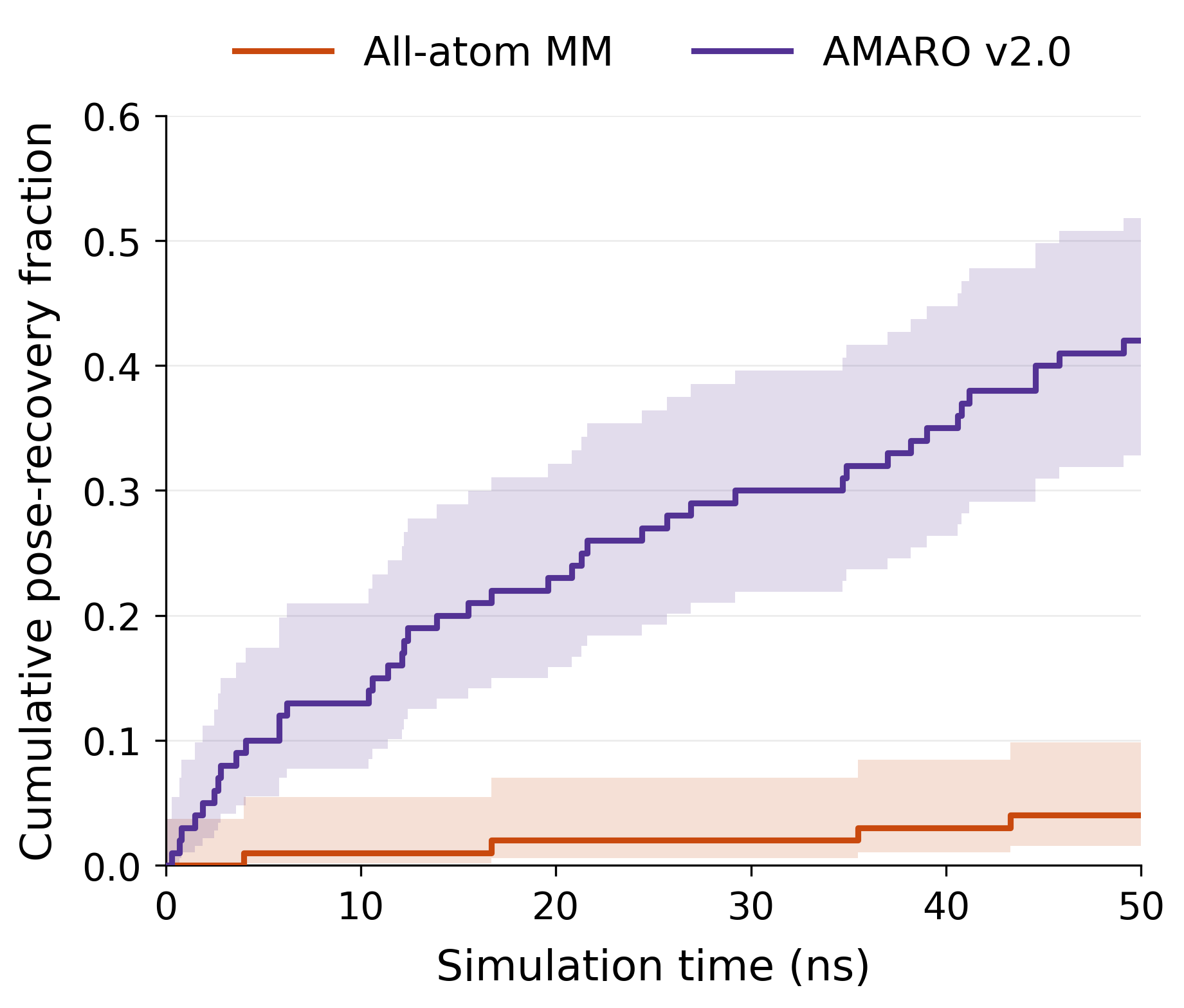}
    \caption{Cumulative first-passage fractions for 100 matched trypsin--benzamidine binding trajectories per model over 50~ns. First passage was defined as the first frame with a ligand RMSD of 2.5~\AA{} or less relative to the 3PTB pose after protein C$\alpha$ alignment. Solid step curves show empirical cumulative fractions, and shaded bands show pointwise 95\% Wilson confidence intervals.}
    \label{sup_fig:tryps_pose_recovery}
\end{figure}

Supplementary Figure~\ref{sup_fig:tryps_pose_recovery} shows empirical cumulative first-passage fractions and pointwise 95\% Wilson confidence intervals. The 50~ns endpoint fractions reported in Results were calculated directly from the observed trajectory outcomes.

\FloatBarrier

\subsection*{Implied timescales of the MSMs}

\subsubsection*{Trypsin--benzamidine}
\begin{figure}[!htbp]
    \centering
    \begin{subfigure}[b]{0.48\textwidth}
        \centering
        \includegraphics[width=\linewidth]{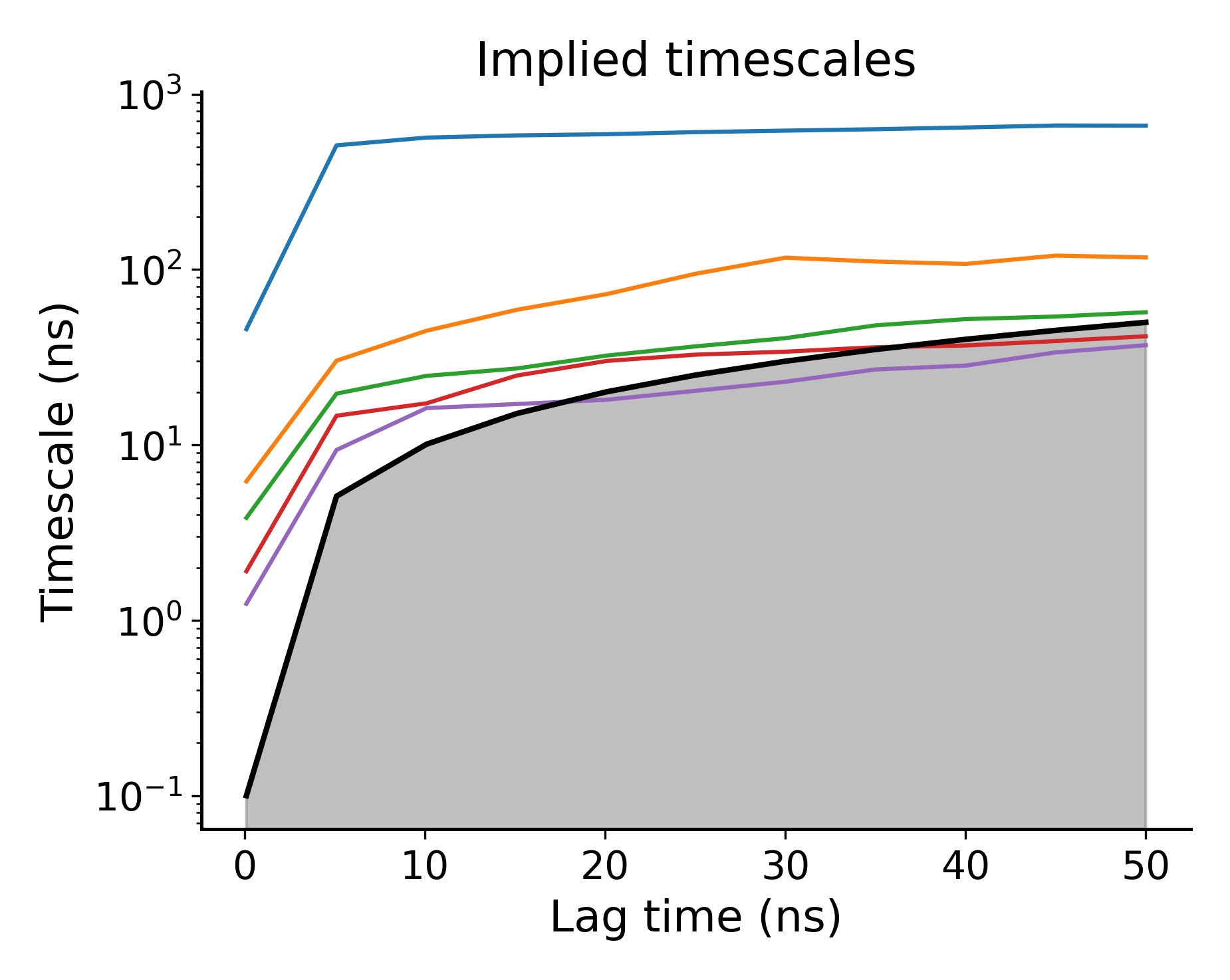}
        \caption{All-atom reference}
    \end{subfigure}
    \hfill
    \begin{subfigure}[b]{0.48\textwidth}
        \centering
        \includegraphics[width=\linewidth]{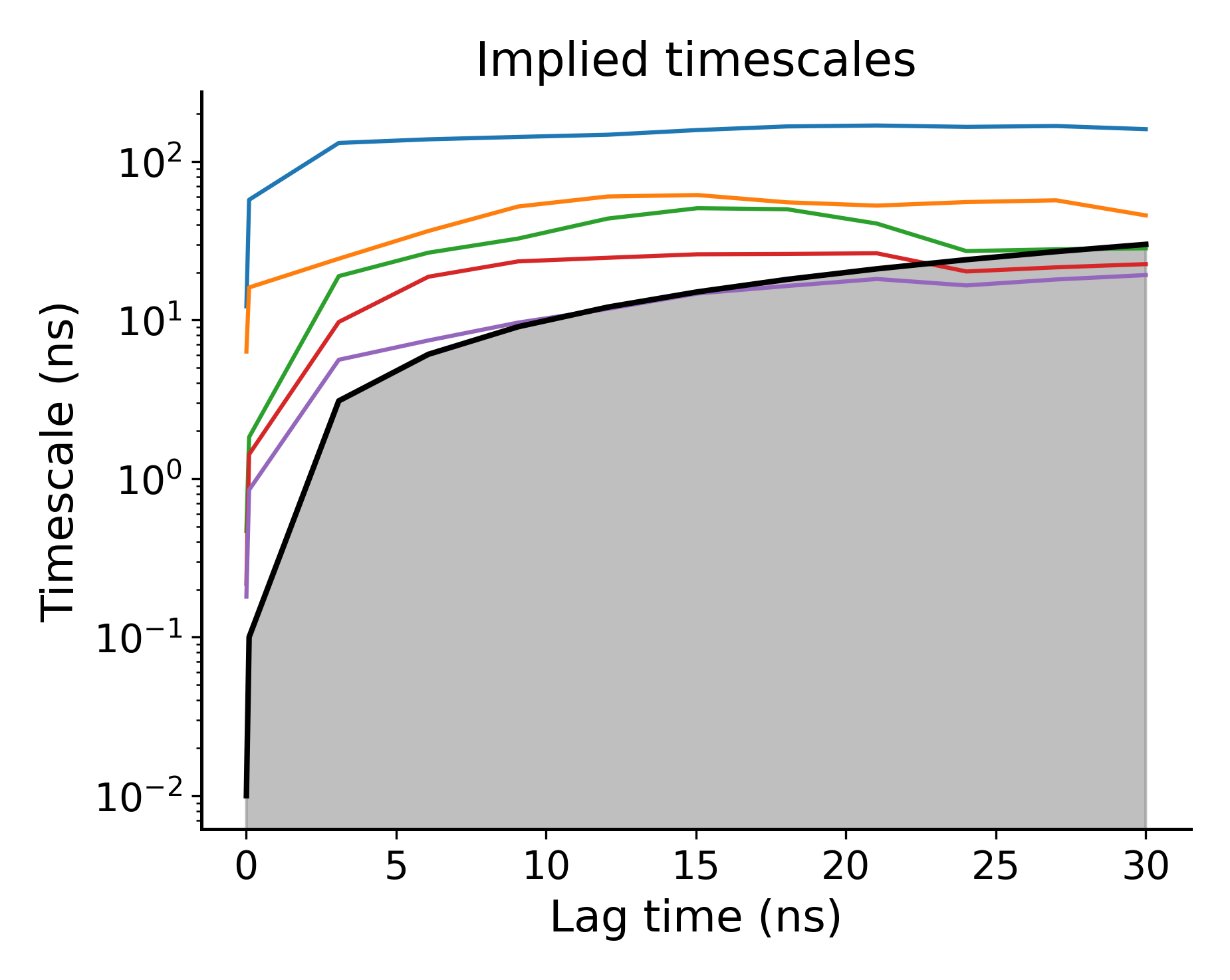}
        \caption{\MODELNAME{}}
    \end{subfigure}
    \caption{Implied timescales of the trypsin--benzamidine MSMs for (a) all-atom MM and (b) \MODELNAME{}. Colored curves show the five slowest relaxation modes as a function of MSM lag time, with timescales plotted on a logarithmic scale. The black line marks equality between timescale and lag time; the gray region indicates timescales shorter than the lag. The reported MSMs use lag times of 20~ns for all-atom MM and 10~ns for \MODELNAME{}.}
    \label{sup_fig:tryps_its}
\end{figure}

Both the all-atom and \MODELNAME{} MSMs show a slowest process separated from the remaining relaxation modes (Supplementary Figure~\ref{sup_fig:tryps_its}). After an initial rise at short lags, the slowest timescale is approximately stable from 10~ns onward, remaining around 0.6~$\mu$s over lag times of 10--50~ns for all-atom MM and 0.15--0.17~$\mu$s over 10--30~ns for \MODELNAME{}.

\FloatBarrier

\subsection*{Protein--ligand structural benchmark}

\subsubsection*{Ligand retention and sequence overlap}

\begin{figure}[!htbp]
    \centering
    \includegraphics[width=0.92\textwidth]{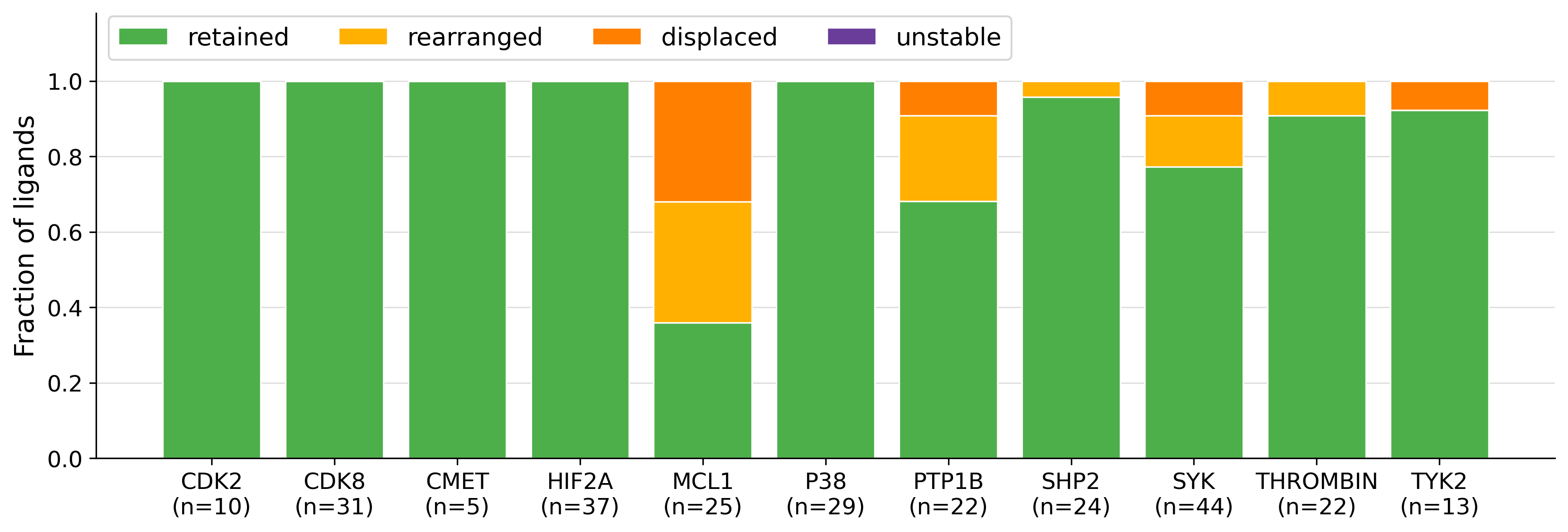}
    \caption{Final ligand outcomes across the OpenFF benchmark. Bars report the fraction of complexes classified as retained (tail-median RMSD $<4$~\AA{}), rearranged (4--6~\AA{}), displaced ($>6$~\AA{}), or unstable. No trajectories were classified as unstable.}
    \label{sup_fig:openff_ligand_outcomes}
\end{figure}

\begin{table}[!htbp]
    \centering
    \caption{Protein and ligand sizes and ligand-retention outcomes across the OpenFF benchmark after 10~ns of \MODELNAME{} simulation. Sizes are expressed as numbers of \textit{noh} beads; outcome columns report numbers of complexes.}
    \label{tab:openff_complex_summary}
    \small
    \begin{tabular}{@{}lrrrrrr@{}}
        \toprule
        Target &
        Protein size &
        Ligand size range &
        Retained &
        Rearranged &
        Displaced &
        Unstable \\
        & (Num. beads) & (Num. beads) & & & & \\
        \midrule
        CDK2     & 4481 & 24--28 & 10 &  0 & 0 & 0 \\
        CDK8     & 5203 & 17--31 & 31 &  0 & 0 & 0 \\
        cMET     & 2276 & 27--37 &  5 &  0 & 0 & 0 \\
        HIF2A    & 1819 & 22--29 & 37 &  0 & 0 & 0 \\
        MCL1     & 1217 & 22--27 &  9 &  8 & 8 & 0 \\
        p38      & 2822 & 23--34 & 29 &  0 & 0 & 0 \\
        PTP1B    & 2431 & 21--37 & 15 &  5 & 2 & 0 \\
        SHP2     & 4216 & 21--25 & 23 &  1 & 0 & 0 \\
        SYK      & 2237 & 25--42 & 34 &  6 & 4 & 0 \\
        Thrombin & 2324 & 26--30 & 20 &  2 & 0 & 0 \\
        TYK2     & 2340 & 21--25 & 12 &  0 & 1 & 0 \\
        \midrule
        Total    &      &        & 225 & 22 & 15 & 0 \\
        \bottomrule
    \end{tabular}
    \vspace{0.5em}
    \begin{minipage}{0.98\textwidth}
        \footnotesize
        Ligand outcomes were classified using the median protein-aligned ligand RMSD over the final 25\% of each trajectory: retained ($<4$~\AA{}), rearranged (4--6~\AA{}), or displaced ($>6$~\AA{}). For heterodimeric systems, alignment used the ligand-binding monomer. We classified trajectories as unstable when ligand RMSD exceeded 50~\AA{} for at least five consecutive analyzed frames, along with protein RMSD above 10~\AA{}.
    \end{minipage}
\end{table}

\FloatBarrier

We assessed receptor overlap using one representative sequence for each OpenFF target. Global alignments to receptors in the protein--ligand training set used BLOSUM62 substitution scores \cite{henikoff1992amino} and affine gap penalties. Sequence identity was defined as the number of identical aligned residues divided by the length of the longer sequence, and each target was matched to the training receptor with the highest identity. Exact matches required identical sequences, whereas near-identical matches required at least 95\% sequence identity and 95\% coverage of both sequences.

\begin{figure}[!htbp]
    \centering
    \includegraphics[width=0.8\textwidth]{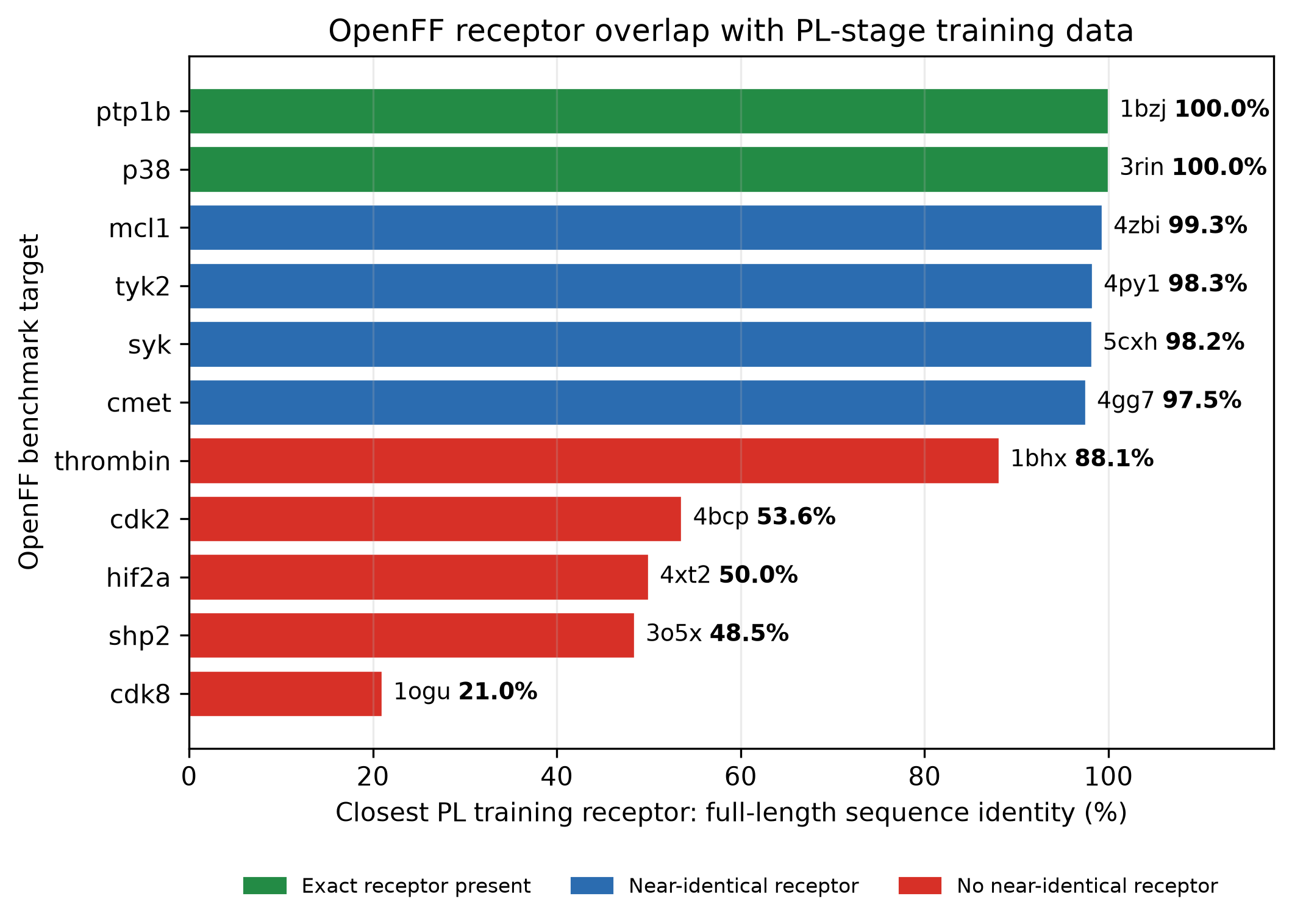}
    \caption{Sequence overlap between OpenFF benchmark receptors and the protein--ligand training set. Bars show full-length sequence identity to the closest training receptor; labels report its PDB identifier and identity. Green denotes exact matches, blue near-identical receptors with at least 95\% identity and coverage, and red the remaining targets. The dashed line marks 95\% identity.}
    \label{sup_fig:openff_receptor_overlap}
\end{figure}

\FloatBarrier

\subsubsection*{Representative MCL1 ligand trajectories}
The MCL1 trajectories illustrate distinct ligand responses within a structurally stable receptor. Ligand 46 remains close to its minimized pose, with fluctuations of about 1--2~\AA{}, whereas ligand 67 shifts by about 3--5~\AA{} via a local torsional rearrangement. Ligand 28 is displaced by more than 100~\AA{} near the end of the trajectory while the protein C$\alpha$-RMSD remains near 1--2~\AA{}, indicating ligand displacement rather than global protein instability.

\begin{figure}[!htbp]
    \centering
    \includegraphics[width=0.8\textwidth]{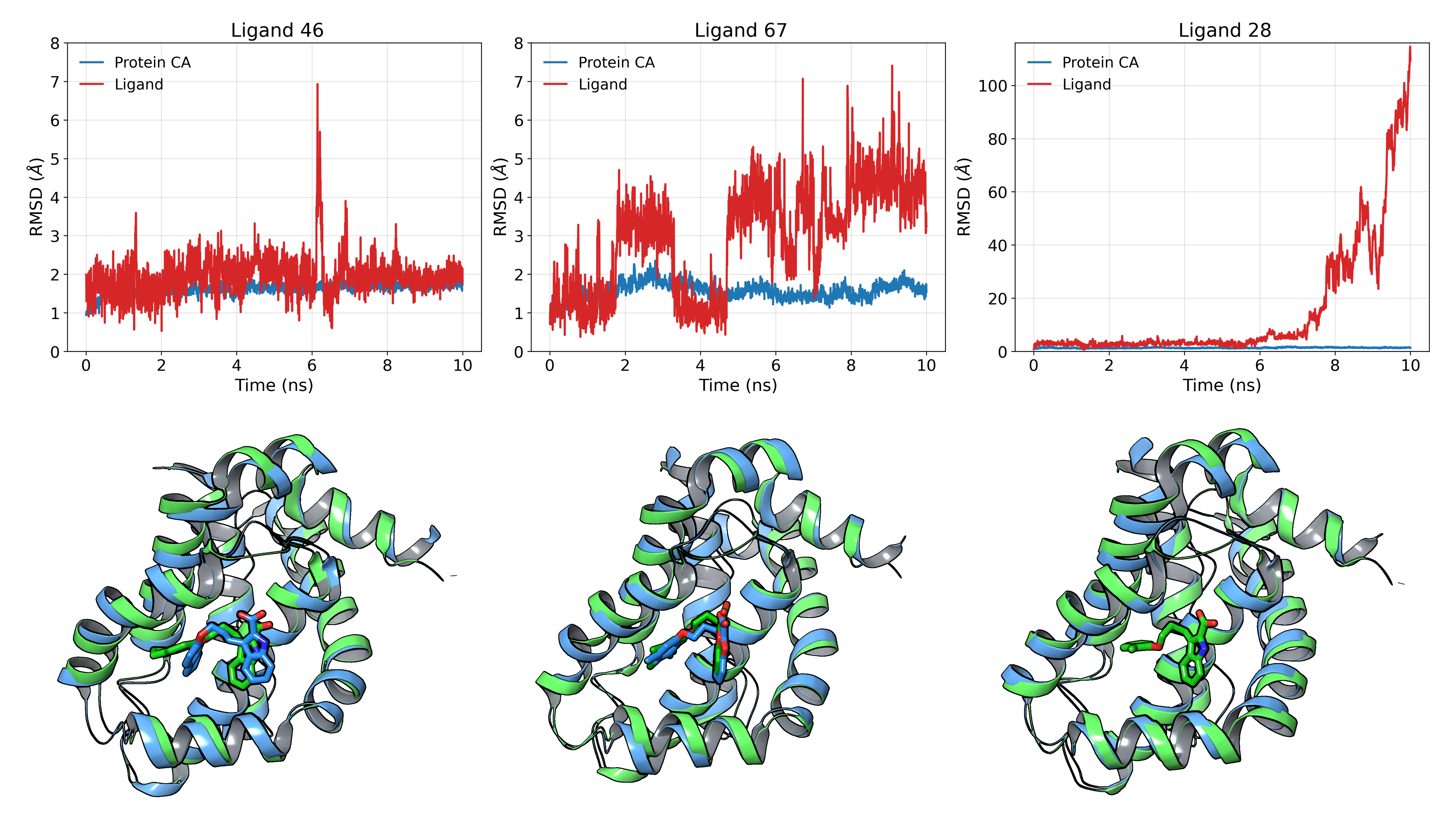}
    \caption{Representative MCL1 ligand-retention outcomes. Top, protein C$\alpha$-RMSD and ligand RMSD for ligands 46, 67, and 28 over 10~ns. Bottom, minimized structures are shown in green and sampled configurations in blue; the final analyzed frame is shown for ligands 46 and 28, whereas the 4~ns frame is shown for ligand 67 to illustrate its torsional rearrangement.}
    \label{sup_fig:mcl1_case_ligands}
\end{figure}

\FloatBarrier

\subsection*{Binding exploration}
\begin{table*}[!htbp]
\centering
\caption{Structural assignment of the ligand-contact macrostates for TYK2--EJM31, SHP2--SHP099, and CDK8--ligand 43. Assignments are based on representative ligand configurations and nearby protein residues shown in Figure~\ref{fig:additional_binding_systems}. Macrostate labels are system-specific.}
\label{sup_tab:additional_binding_macrostates}
\begin{tabularx}{\textwidth}{ll l X}
\toprule
System & Macrostate & Representative region & Structural assignment \\
\midrule

TYK2--EJM31
& M0 & Val981
& ATP cleft; overlaps the reference binding region~\cite{liang2013tyk2}  \\

& M1 & Thr890
& TYK2 surface contact \\

& M2 & Tyr1055
& TYK2 surface contact \\

& M3 & Val1121
& TYK2 surface contact \\

& M4 & Lys1139
& TYK2 surface contact \\

\midrule

SHP2--SHP099
& M0 & Tyr80/Lys280
& N-SH2--PTP latch site~\cite{fodor2018dual} \\

& M1 & Leu436
& Exposed PTP surface contact \\

& M2 & His114/Thr253
& Canonical SHP099 allosteric tunnel~\cite{chen2016shp2} \\

& M3 & Tyr197/Glu204
& Exposed C-SH2 surface contact \\

\midrule

CDK8--ligand 43
& M0 & Pro124
& Distal CDK8 surface contact \\

& M1 & CDK8 Trp146/Cyclin C Arg58
& Off-site contact spanning the CDK8--Cyclin C region \\

& M2 & Cyclin C Asp182
& Cyclin C surface contact \\

\bottomrule
\end{tabularx}
\end{table*}

\begin{figure}
  \centering
  \includegraphics[width=0.85\linewidth]{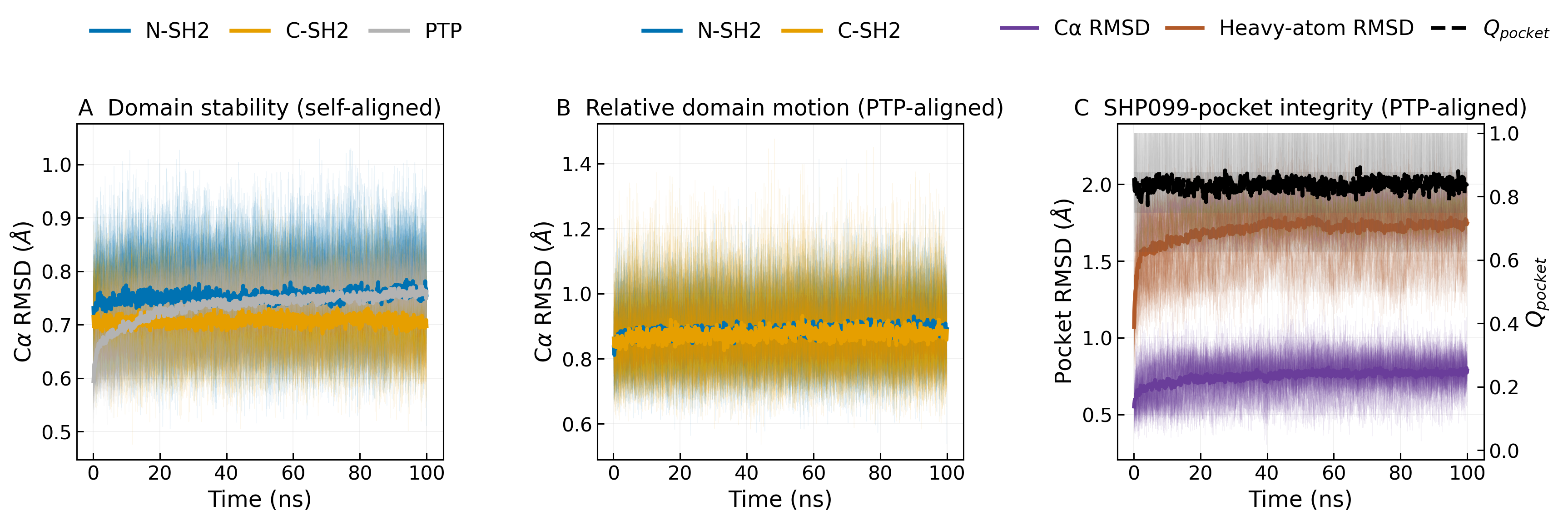}
  \caption{SHP2 domain and pocket geometry during restrained binding-exploration simulations. Protein C$\alpha$ atoms more than 9~\AA{} from the reference ligand were restrained as described in Methods. (a) Self-aligned C$\alpha$-RMSD of the N-SH2, C-SH2, and PTP domains relative to the initial structure. (b) N-SH2 and C-SH2 RMSDs after PTP alignment, showing relative domain motion. (c) PTP-aligned SHP099-pocket C$\alpha$ and heavy-atom RMSDs (left axis), and the fraction of reference interdomain pocket contacts, $Q_{\mathrm{pocket}}$ (right axis). Thin traces show individual trajectories and thick curves show averages across 50 independent 100~ns simulations. $Q_{\mathrm{pocket}}$ measures retention of eight reference residue--residue contacts, defined by a minimum inter-residue heavy-atom distance of 4.5~\AA{} or less.}
  \label{sup_fig:shp2_rearrangement}
\end{figure}

\FloatBarrier
\subsection*{Trypsin--benzamidine macrostate composition}

\begin{figure}[!htbp]
    \centering
    \begin{subfigure}[b]{0.9\textwidth}
        \centering
        \includegraphics[width=0.8\linewidth]{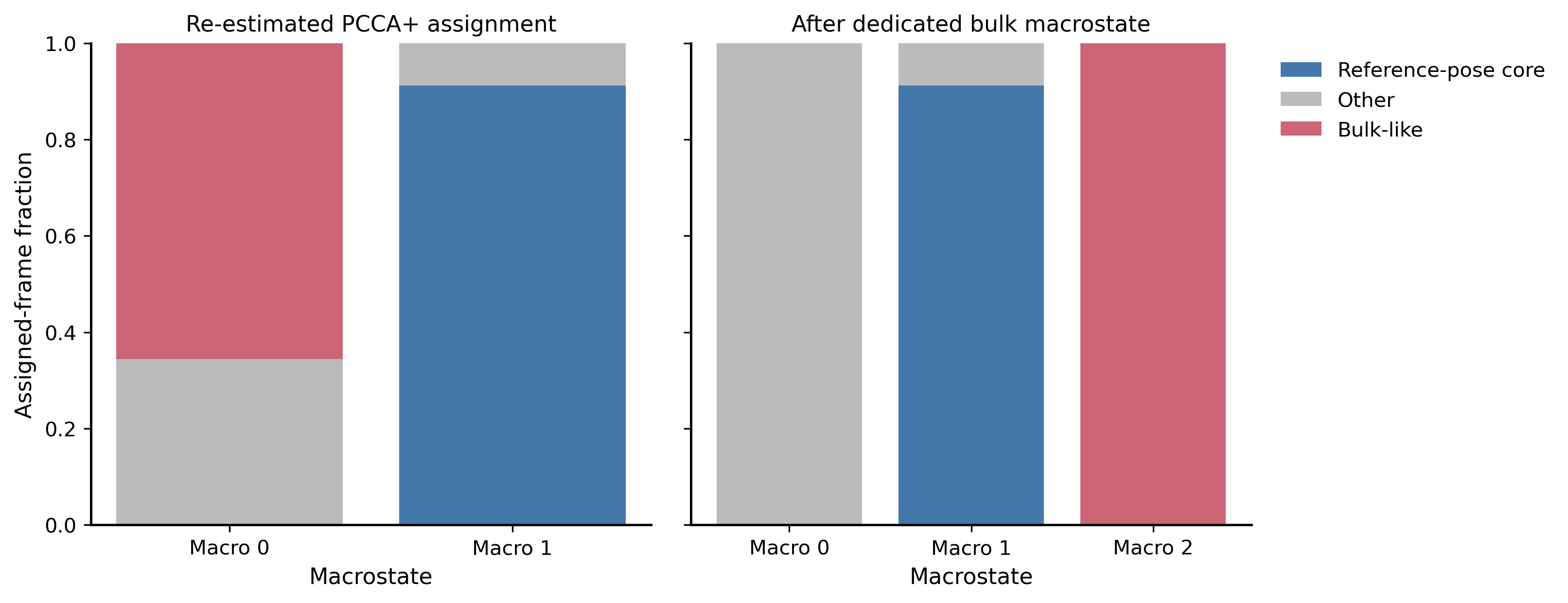}
        \caption{All-atom MM}
    \end{subfigure}
    \par\medskip
    \begin{subfigure}[b]{0.8\textwidth}
        \centering
        \includegraphics[width=0.9\linewidth]{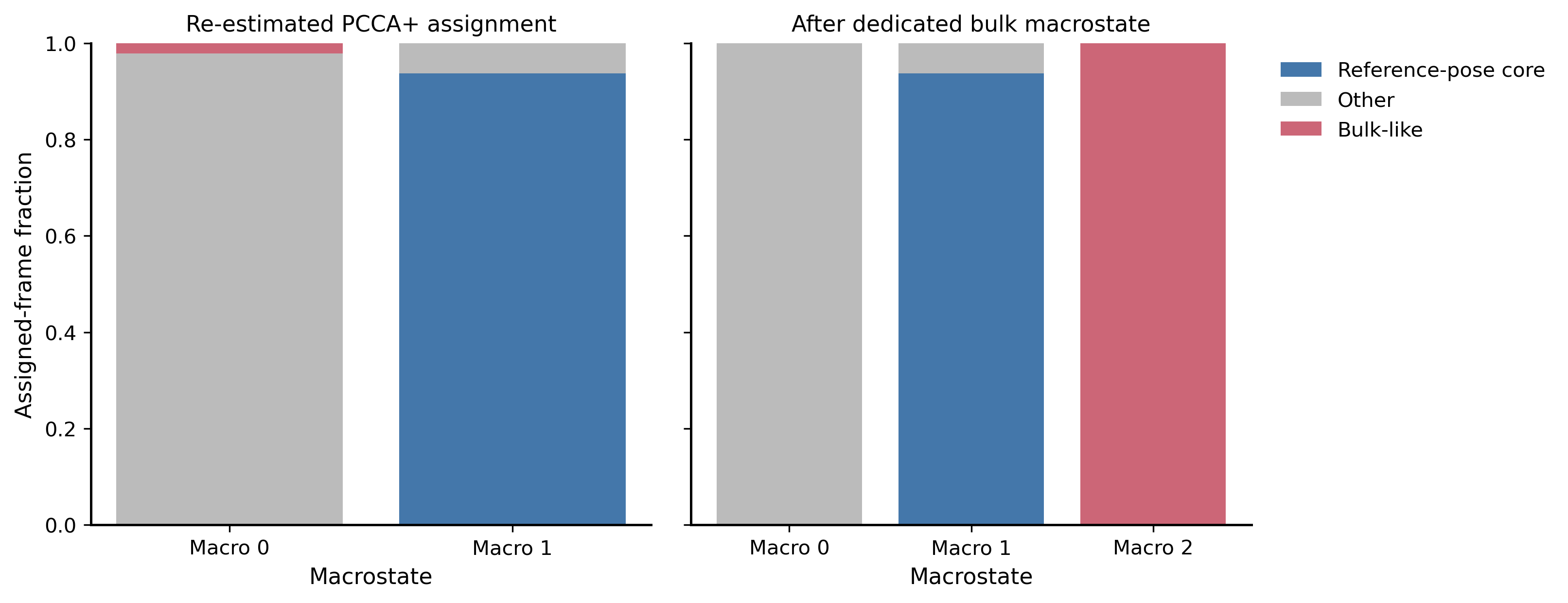}
        \caption{\MODELNAME{}}
    \end{subfigure}
    \caption{Trypsin--benzamidine macrostate composition before and after dedicated bulk-state grouping at the 12~\AA{} cutoff for (a) all-atom MM and (b) \MODELNAME{}. Left, the two-state PCCA+ partition fitted after splitting mixed bulk/nonbulk microstates. Right, the final three-macrostate partition after grouping wholly bulk microstates into M2, with the fitted microstate transition matrix unchanged. Bars report fractions of assigned frames within each macrostate: reference-pose core (blue; protein-aligned ligand RMSD $\leq2.5$~\AA{}), bulk-like frames (pink; minimum ligand-COM--residue-COM distance $>12$~\AA{}), and other frames (gray). M1 is the reference-pose-rich bound sink, M2 the dedicated bulk source, and M0 the remaining nonbulk state.}
    \label{sup_fig:tryps_macrostate_composition}
\end{figure}

\FloatBarrier
\subsection*{Computational performance}

\MODELNAME{} contains 236,000 trainable parameters, compared with 576,000 for AMARO v1.0. On an NVIDIA RTX 4090 GPU with a 4~fs time step, the compiled \MODELNAME{} implementation was about fourfold faster than AMARO v1.0 across the five systems benchmarked for both model generations (Table~\ref{sup_tab:computational_benchmark}). The Acellera-TensorNet plugin implementation further increased throughput, providing a 15--22-fold speedup over AMARO v1.0 and enabling simulations of systems containing up to 30,000 \textit{noh} sites on a single GPU.

\begin{table*}[!hbtp]
    \centering
    \small
    \caption{Molecular dynamics benchmark on an NVIDIA RTX 4090 GPU using a 4~fs time step. Site counts correspond to the complete \MODELNAME{} representation. For AMARO v1.0, we excluded two unsupported sites from TYK2 and SHP2 because the corresponding element-based bead types are absent from its \textit{noh} representation, resulting in 2,359 and 4,236 evaluated sites, respectively. Dashes indicate configurations that exceeded the available GPU memory.}
    \label{sup_tab:computational_benchmark}
    \begin{tabular}{@{}lrrrr@{}}
    \toprule
    \textbf{System} & \textbf{\textit{noh} sites} & \multicolumn{3}{c}{\textbf{Throughput [ns day$^{-1}$]}} \\
    \cmidrule(l){3-5}
     & & \textbf{AMARO v1.0} & \textbf{\MODELNAME{} (compiled)} & \textbf{\MODELNAME{} (Acellera-TensorNet)} \\
    \midrule
    1UBQ & 602 & 16.7 & 67.6 & 249 \\
    MCL1 & 1,239 & 9.81 & 38.9 & 203 \\
    TYK2 & 2,361 & 5.19 & 20.5 & 116 \\
    SHP2 & 4,238 & 2.76 & 11.7 & 60.3 \\
    CDK8 & 5,229 & 2.33 & 9.39 & 47.7 \\
    1XNJ & 9,306 & -- & 4.96 & 24.2 \\
    7RAS & 20,000 & -- & 2.28 & 26.2 \\
    4AV7 & 29,682 & -- & -- & 6.96 \\
    \bottomrule
    \end{tabular}
\end{table*}

\end{document}